\documentclass[11pt]{article}%
\usepackage[letterpaper, left=1in, top=1in, right=1in, bottom=1in,nohead,includefoot]{geometry}
\usepackage{charter} 
\usepackage{graphicx,svgcolor} 
\usepackage[svgnames,x11names]{xcolor}
\usepackage[round]{natbib}
\usepackage{url,bm,amsmath,amsfonts,amssymb,setspace,subfigure,placeins} 
        \usepackage[compact]{titlesec} 
        \titleformat*{\section}{\normalfont\Large\bfseries\blu}
        \titleformat*{\subsection}{\normalfont\large\bfseries\blu}
        \titleformat*{\subsubsection}{\normalfont\normalsize\bfseries\blu}
        \usepackage[pdftex]{hyperref}
        \hypersetup{colorlinks,%
	           linkcolor=MidnightBlue,  
	           urlcolor=DarkSlateGray,   
	           citecolor=RoyalBlue3  
	   }
 
\def\blu{\color{RoyalBlue4}}         
\def\bone{\mathbf{1}}
\def\bzero{\mathbf{0}} 
\def\A{\mathbf{A}} 
 
\def\I{\mathbf{I}} 

\def\V{\mathbf{V}}
\def\W{\W}  
\def\a{\textrm{a}}
\def\c{\textrm{c}}
\def\e{\textrm{e}}
  
\def\g{\mathbf{g}}  

\def\m{\mathbf{m}}
\def\r{\mathbf{r}}
\def\s{\mathbf{s}}  
\def\w{\mathbf{w}}
\def\x{\mathbf{x}}
\def\y{\mathbf{y}}
\def\z{\mathbf{z}} 
 
\def\H{\mathbf{H}}
\def\M{\mathbf{M}}
\def\R{\mathbf{R}}  
\def\C{\mathbf{C}} 
\def\V{\mathbf{V}} 
\def\W{\mathbf{W}} 

\def\bepsilon{{\bm\epsilon}}
\def\bvarepsilon{{\bm\varepsilon}}

\def\bPsi{{\bm\Psi}}

\def\bOmega{{\bm\Omega}}

\def\btau{{\bm\tau}}

\def\cM{{\mathcal M}}

\def\seq#1#2{#1{:}#2}
\def\eqn#1{eqn.~(\ref{eq:#1})}

\def\bi{\begin{itemize}[noitemsep,topsep=3pt]}
\def\ei{\end{itemize}}
\def\bn{\begin{enumerate}[noitemsep,topsep=3pt]}
\def\bc{\begin{center}}\def\ec{\end{center}}
\def\en{\end{enumerate}}
\def\beq#1{\begin{equation}\label{eq:#1}}\def\eeq{\end{equation}}
\def\beas{\begin{eqnarray*}}\def\eeas{\end{eqnarray*}}
\def\bea{\begin{eqnarray}}\def\eea{\end{eqnarray}}

\def\RHmp{(\R\H)^+}\def\RHmpt{(\R\H)^{+'}}  

\def\cM{{\mathcal M}}

\usepackage[font=small,skip=5pt]{caption}

\begin{document}\emergencystretch 3em  

\title{\bf\blu   Decision Synthesis in Monetary Policy}

\author{Tony Chernis\\
	{\small Bank of Canada} \\
	\and Gary Koop \\
	{\small University of Strathclyde} \\
        \and Emily Tallman\\
        {\small Duke University}
\and Mike West\\
	{\small Duke University}
}

\maketitle
\begin{center}
\begin{minipage}{0.95\textwidth}

\bc{\bf \blu Abstract}\ec
The macroeconomy is a sophisticated dynamic system involving significant uncertainties that complicate modelling. In response, decision-makers consider multiple models that provide different predictions and policy recommendations which are then synthesized into a policy decision. In this setting, we develop Bayesian predictive decision synthesis (BPDS) to formalize monetary policy decision processes. BPDS draws on recent developments in model combination and statistical decision theory that yield new opportunities in combining multiple models, emphasizing the integration of decision goals, expectations  and outcomes into the model synthesis process.  Our case study concerns central bank policy decisions about target interest rates with a focus on implications for multi-step macroeconomic forecasting. This application also motivates new methodological developments in conditional forecasting and BPDS, presented and developed here.


\medskip
{\em Keywords}: Bayesian predictive decision synthesis, dynamic model averaging, interest rates, macroeconomic forecasting, model combination, path forecasting, policy decisions, scenario analysis
\end{minipage}
\end{center}

\thispagestyle{empty} 
\newpage


\setcounter{page}{1}
\section{Introduction}

Monetary policymakers are tasked with simple, but hard to achieve, objectives. A key example is the \lq\lq dual mandate" under which  policymakers target future inflation rates and real activity using interest rates as the policy instrument. Decisions are made based on uncertain information from many sources. Here, the sources are models that generate predictive distributions for macroeconomic outcomes and policy instruments over multiple time periods. For a single model, an optimal policy path is evaluated via conditional forecasting and decision analysis. Applying standard methods, such as Bayesian model averaging (BMA), is one way to address model uncertainty-- routine decision analysis can then be applied to the weighted average of models. This traditional view, however, ignores the reality that models may each individually recommend very different optimal policy decisions. The question then arises of how to synthesize this information and, potentially, exploit it in the overall final decision process. This paper addresses this question. 

The extensive Bayesian econometrics literature on model combination rarely highlights the fact that models are typically built for specific prediction and decision goals. Traditional BMA analysis weights models according to purely statistical model fit and only scores one-step ahead forecast outcomes. Extensions and alternatives have arisen to define model weightings based on aspects of past forecast performance with respect to specific forecast goals. \cite{MARTIN2023} survey Bayesian forecasting in economics and finance and review various forecast combination approaches, including some that are
more explicitly concerned with goal-focused prediction~\citep[e.g.,][]{mitchell-evaluating-2005, geweke-optimal-2011, conflitti-optimal-2015,Kapetanios2015,LoaizaMayaJOE2021,chernis-nowcasting-2022,AastveitEtAL2022, BernaciakGriffin2024}.  
\cite{LavineLindonWest2021avs} place many of the earlier approaches in a foundational Bayesian context. They justify model weights based on utilities in forecasting using historical model-specific \lq\lq scoring\rq\rq\ of past forecast outcomes. The underlying theoretical justifications come from Bayesian predictive synthesis (BPS) and the specific class of \lq\lq mixture BPS" models~(\citealp{McAlinnWest2018}, section 2.2;~\citealp{JohnsonWest2022}). However, while ultimate decision goals may be implicit in specific applications of model combination, they are rarely, if ever, taken into account in the analysis and resulting decision-making. This raises concerns. A model that has fit or forecast specific outcomes well in the past may be a good bet for use in resulting decision analysis, such as defining optimal decisions about values of policy instruments; however, there is no guarantee that this will be so.

Our view is that models that have recommended policy decisions that turned out to be \lq\lq good" should be more heavily weighted in looking ahead, just as past statistical predictive performance is generally positively weighted. 
The challenge is to operationalize the concept of \lq\lq good decision" performance. For example, a vector autoregression (VAR) model can be evaluated on forecast performance using a pseudo real-time forecasting exercise, but it is not clear how to evaluate such a model when it is used to advise policy decisions. We can, however,  explore how the model would have advised on decisions in the past. Evaluations can then compare such analyses to decisions actually made by policymakers in the past (while recognizing that policymakers' past decisions were not necessarily \lq\lq good" and just outcomes of the rather amorphous reality of monetary policymaking). 
 
Bayesian predictive decision synthesis (BPDS-- \citealp{TallmanWest2023}) has the potential to address these questions. An outgrowth of the theoretical BPS framework, BPDS explicitly addresses model scoring based on decision outcomes as well as predictive accuracy. In addition to reflecting historical outcomes of predictions and decisions, BPDS allows for differential model weighting based on~\textit{expected} decision outcomes. This is a decision parallel to the proven use of BPS models that incorporate outcome-dependent weights that modify BMA-like mixtures to differentially favour models in different parts of the future outcome space for pure forecasting. The latter concept was introduced by~\cite{Kapetanios2015} whose empirically inspired developments recognized, for example, that one model may be better at predicting inflation when inflation is high and rising, while another model may be better when inflation is low and stable. BPS defines a theoretical Bayesian basis and broader methodological framework for this~\citep{JohnsonWest2022}. BPDS goes further by integrating both historical and expected decision outcomes; here we develop, extend, and exemplify BPDS in the central macroeconomic policy context. 

BPDS applies the Bayesian mixture model approach of BPS using defined utility-- or \lq\lq score"-- functions that relate to explicit decision goals. 
This allows for multiple objectives (i.e., multi-attribute decision analysis). For example, a purely predictive vector score function can allow for multiple forecast horizons (e.g., to produce inflation near a target for each of the next eight quarters) and/or multiple outcome criteria (e.g., to separately reflect inflation targeting, interest rate smoothing, and stable growth patterns over coming quarters), among others. For policymakers juggling multiple objectives, this key feature is rather distinct from conventional approaches that adopt single, scalar criteria for model weighting. For example, a forecast combination approach might choose model weights based on the $h-$step ahead predictive likelihood for a single choice of $h$, with BMA simply focused on $h=1$. In contrast, BPDS can address multi-steps ahead in parallel, along with scoring of realized decisions that simultaneously target several macroeconomic outcomes.  

The opportunities for exploring such practically relevant questions are highlighted in empirical studies. Our case study uses US macroeconomic data with multi-objective score functions to define BPDS model weights, demonstrating  BPDS in macroeconomic forecasting and advisory decision-making.   This is enabled by new methodological advances motivated by this applied context.  First, in developing the scenario-based conditional forecasting analyses, we explicitly recognize the dual roles of policy decision variables (such as interest rates) as both outcomes and putative controls. This leads to an important theoretical extension of traditional conditional forecasting in which candidate decision scenarios are weighted by their predicted plausibility prior to assessing their likely implications in forecasting outcomes of other variables.\footnote{This is keeping in the spirit of \cite{LEEPER2003} who argue that the \lq\lq modest interventions"commonly used in reduced-form scenario analysis are unlikely to be subject to the Lucas Critique. We explicitly give higher weight to small interventions and down-weight large interventions.} This is important quite generally in conditional forecasting, as well as here in the resulting BPDS analysis. Second, we advance the BPDS methodology based on a new and practically critical focus on the relationships between multiple utilities defining the scoring of decision outcomes. Third, we discuss and promote the use of bounded utilities in model scoring, and customized forms of utility functions to guide model-specific and final BPDS-based optimal decisions.

\section{BPDS Framework} 

We present and discuss the structure of BPDS at a particular point in time, ignoring the time dependency and relevance in the notation for clarity in communicating these essentials. Practical implementation in time series is of course sequential, with models at time $t$ depending on all relevant historical data and information. 

\subsection{Mixture BPDS and Decision Setting}  

At a given time point, let $\y$ denote 
the $q-$dimensional outcome variable of interest (e.g., inflation in each of the next $q$ quarters) and $\x$ the vector of control/decision variables (e.g., a target profile of central bank interest/base rates over the next $q$ quarters). Each of a set of $J$ models, $\cM_j$, $j=\seq1J,$ predicts the outcome $\y$ via a predictive density $p_j(\y| \x, \cM_j)$ conditional on any considered decision $\x.$ The policymaker responsible for ultimate decisions adopts a general BPDS approach with the overall conditional (on $\x$) predictive p.d.f.  
\begin{equation} \label{eq:BPDSmixgeneral} 
f(\y|\x) \propto \sum_{j=\seq 0J} \pi_j(\x) \alpha_j(\y|\x) p_j(\y|\x,\cM_j) 
\end{equation}
with the following ingredients.

\subsubsection*{\em BPDS model probabilities} 

The {\em decision-dependent model probabilities} $\pi_j(\x)$ can differentially weight models $j$ over the decision space $\x$. This incorporates any prior information relevant to model weighting based on past predictive model fit and decision outcomes, and now explicitly allows for adjustments based on a currently considered decision $\x$. Dependence of $\pi_j(\x)$ on $\x$ is simply fundamental and critical in our policy setting.

\subsubsection*{\em BPDS calibration functions} 

The $\alpha_j(\y|\x)$ are {\em calibration functions} that define outcome dependence of model weights over the outcome space of $\y$ for any chosen $\x$. This defines the opportunity to increase or decrease the model weights differentially over the outcome $\y$ space to address model-specific biases and preferences and address questions of model-specific calibration more generally. The BPDS mixture of~\eqn{BPDSmixgeneral} has the equivalent form
\begin{equation} \label{eq:BPDSmixgeneralfform} 
f(\y|\x) = \sum_{j=\seq 0J} \tilde\pi_j(\x) f_j(\y|\x,\cM_j) 
\end{equation}
where
\begin{equation}\label{eq:fj}
f_j(\y|\x,\cM_j)=\alpha_j(\y|\x) p_j(\y|\x,\cM_j) /a_j(\x) \quad\textrm{and}\quad \tilde{\pi}_j(\x)=k(\x) \pi_j(\x) a_j(\x) 
\end{equation}
with normalizing terms $k(\x)$ and $a_j(\x)$ explicitly dependent on $\x$. This form shows how the calibration functions $\alpha_j(\cdot|\cdot)$ modify the initial mixture pdfs $p_j(\cdot|\cdot) \to f_j(\cdot|\cdot)$ with corresponding changes of mixture weights $\pi_j(\x)\to\tilde\pi_j(\x).$ 

Note that the choice of relevant calibration functions $\alpha_j(\y|\x)$ will, in any given application, be partly dependent on characteristics of the model pdfs $p_j(\y|\x,\cM_j)$. In particular, the expectation of each $\alpha_j(\y|\x)$ under $p_j(\y|\x,\cM_j)$ must be finite in order that~\eqn{BPDSmixgeneral} defines a valid BPDS density $f(\y|\x).$ 
This supports the use of bounded scores, in general.

\subsubsection*{\em Baseline mixture component} 

The model index $j=0$ explicitly allows for a {\em baseline} model component $\cM_0$ in the mixture p.d.f. $f(\cdot|\cdot)$ that can, among other things, address the ever-present issue of \lq\lq model set incompleteness"~\citep[][section 2.2.3]{TallmanWest2023}. $\cM_0$ can be chosen to produce a pdf
$f_0(\cdot|\cdot)$ that is over-dispersed relative to the mixture of the initial $J$ models, so supporting outcomes $\y$ that are unusual under the $J$ models. 
The baseline is then a suitable \lq\lq fall back" model for times when the other models are forecasting poorly.

\subsubsection*{\em Initial mixture} 

The special case with each $\alpha_j(\y|\x)=1$ defines the {\em initial mixture} with no BPDS calibration. We use $p(\y|\x)$ in notion; that is, $p(\y|\x) =  \sum_{j=\seq 1J} \pi_j(\x) p_j(\y|\x,\cM_j)$.

Special cases fix ideas. First, if $\pi_j(\x)=\pi_j$ with $\pi_0=0$ are model probabilities based on historical BMA analysis, and with $\alpha_j(\y|\x)=1,$ then~\eqn{BPDSmixgeneral} specializes to BMA. Thus, BMA analyses-- with or without this decision dependence in model-specific forecasts-- are very special cases of BPDS. Second, again with $\pi_j(\x)=\pi_j$, $\pi_0=0$ and $\alpha_j(\y|\x)=1,$ the decision-maker has the freedom to specify the initial mixture probabilities $\pi_j$ in other ways than with BMA. This includes using historical performance defined by scoring of past forecast outcomes, justifying various approaches to goal-focused model weighting~\citep[e.g.,][and references therein]{LavineLindonWest2021avs,LoaizaMayaJOE2021} as special cases of BPDS.  
Third, mixture BPS~\citep{McAlinnWest2018,JohnsonWest2022} is a special case in which models are combined with outcome-dependent weights. In these settings, $\pi_j(\x)=\pi_j$ depends on past predictive performance, $p_j(\y|\x) = p_j(\y)$ and $\alpha_j(\y|\x)=\alpha_j(\y)$ define outcome-dependent modifications of model probabilities, but there is no decision context so no $\x-$dependence. BPDS critically recognizes that the foundational BPS theory allows explicit incorporation of decision goals-- admitting the conditioning on $\x$ throughout all components of~\eqn{BPDSmixgeneral}-- to extend the foregoing analyses. 

With predictions of $(\y|\x)$ based on~\eqn{BPDSmixgeneralfform}, the Bayesian decision-maker acts to identify the optimal decision $\x$ based on a chosen utility function $U(\y,\x).$ This involves numerical optimization to maximize the implied expected utility $\bar U(\x) = E_f[U(\y,\x)|\x]$ over the decision space of $\x.$ The notation $E_f[\cdot|\cdot]$ here explicitly represents expectation with respect to the BPDS distribution, and we use $E_p[\cdot|\cdot]$ to denote expectation under the initial mixture. 

\subsection{Decision-dependent Scores for Calibration Functions}\label{sec:DDS}

The key step in integrating decision outcomes into relative model weights addresses the question of how each $\cM_j$ would inform decisions if used alone. 
Given the predictive p.d.f. $p_j(\y|\x,\cM_j)$ and a chosen, potentially model-specific utility function $u_j(\y,\x),$ acting based only on $\cM_j$ leads to the optimal decision $\x_j$ that maximizes $E_{p_j}[ u_j(\y,\x)|\x]$ over $\x.$ The decision-maker has access to this set of model recommendations and is interested in model combination to preferentially weight \lq\lq good decision models" as well as models that generate good predictions. BPDS formalizes this with specified {\em score functions} $\s_j(\y,\x_j),$ each being a $k-$vector of utilities that can be chosen to reflect both predictive and decision goals. The use of multi-dimensional scores addresses multiple goals simultaneously. 

The Bayesian decision-theoretic development of~\cite{TallmanWest2023} generates the resulting functional forms of the BPDS calibration functions as 
\begin{equation} \label{eq:alphajBPDS} 
\alpha_j(\y,\x) = \exp\{ \btau(\x)'\s_j(\y,\x_j)\}, \,\, j=\seq0J,
\end{equation}
where $\btau(\x)$ is a $k-$vector with elements differentially weighting the multiple utility dimensions of the score vector.   The reasoning and theory behind this key result is as follows. 

The initial mixture $p(\y|\x)= \sum_{j=\seq 1J} \pi_j(\x) p_j(\y|\x,\cM_j)$ is the $\y-$margin of the joint distribution 
$p(\y,\cM_j|\x) = \pi_j(\x) p_j(\y|\x,\cM_j),$ $(j=\seq 1J)$. Under this initial distribution for any candidate decision $\x$,  
and with score vectors $\s_j(\y,\x_j)$ defined and evaluated at model-specific optimal decisions $\x_j,$ the decision-maker has {\em initial expected score}  
$\m_p(\x) = \sum_{j=\seq0J} \pi_j(\x)\m_{jp}(\x)$ where 
$\m_{jp}(\x) = \int_{\y} \s_j(\y,\x_j)p_j(\y|\x,\cM_j)d\y.$    Treating $\m_p(\x)$ as a benchmark to improve on expectation, the BPDS theory enquires about 
distributions $f(\y,\cM_j|\x)$ that yield expected scores $\m_f(\x) \ge \m_p(\x) + \bepsilon(\x)$ for some non-negative $k-$vector (with at least one positive entry) $\bepsilon(\x);$ this may be chosen to depend on $\x,$ or may be a specified constant \lq\lq decision score improvement." Given $\m_f(\x),$ the BPDS theory 
identifies a unique $f(\cdot,\cdot|\x)$ that minimizes the K\"ullback-Leibler (KL) divergence of $p(\cdot,\cdot|\x)$ from $f(\cdot,\cdot|\x)$ and has an expected score of exactly $\m_f(\x).$ The theory is that of {\em relaxed entropy tilting}~\citep{TallmanWestET2022, TallmanWest2023,West2023constrainedforecasting} and yields
$f(\y,\cM_j|\x) \propto \pi_j(\x) \alpha_j(\y,\x) p_j(\y|\x,\cM_j)$ with calibration function precisely as in~\eqn{alphajBPDS}. The {\em tilting vector} $\btau(x)$ is implicitly defined by the vector of $k$ {\em target score constraints} $E_f[ \s_j(\y,\x_j) |\x ] = \m_f(\x).$

BPDS uses the initial mixture based on past performance, but then admits that small perturbations of model probabilities based on their expected performance may lead to improved scores. 
A stylized example has $J=2$ models that have forecast equally well to date. Traditional model averaging methods including BMA confer equal weights in the combination. If, however, the models have different expected scores $\m_{jp}(\x)$, conferring slightly more weight on the model expected to lead to a higher score makes sense.   

The entropic (or exponential) tilting theory is general. It is, of course, possible to tilt the initial joint distribution to most targets, so long as they are technically achievable under the initial distribution. However, an overly ambitious target score may result in empirically unreasonable results. Hence, we emphasize the importance of selecting $\m_f(\x)$ that represents a \lq\lq small" improvement over the initial benchmark score $\m_p(\x).$ This is bolstered by the assumption that the initial model probabilities reflect the empirical plausibility of models, as well as any available information about historical predictive and decision performance.  Further, as we demonstrate later in the case study, aspects of the computational methodology for model fitting in the sequential time series setting naturally inform on, and allow monitoring of, relevant choices of target expected scores. 
 
\subsection{BPDS Summary}

This section has outlined the main ideas underlying BPDS and the key ingredients of the theory and resulting technical machinery.
Specifications of score and utility functions, initial model probabilities, and target scores are all required for implementation and are, of course, application specific. The following section develops full details in the context of the macroeconomic decision-making application. In terms of computation, BPDS requires the use of posterior simulation methods (i.e., draws from conditional predictive densities from each model), as well as numerical optimization methods (i.e., to find the $\x_j$ and the overall optimal decision $\x$ from BPDS analysis).

\section{BPDS for Optimal Monetary Policy Decisions\label{sec:BPDS}}

Following~\cite{frs2019}, we use quarterly macroeconomic and financial data from 1973:Q1 to 2022:Q2 from the FRED-QD database (Federal Reserve Bank of St.~Louis). This includes GDP (log of real GDP), prices (log of GDP deflator), interest rate (the shadow rate,\footnote{This is the Federal Funds rate when the latter is positive, but can go negative when it is at the zero lower bound, taking into account unconventional monetary policy; see \cite{Wu-shadowrate}.} which we treat as the policy rate), investment (ratio of real gross private domestic investment to GDP), real stock prices (log of the S\&P500 deflated by GNP/GDP price index), and spread (between BAA bonds and the Fed funds rate). Models are run over multiple years. Each quarter they produce forecasts-- full predictive distributions in terms of Monte Carlo samples-- of outcomes of interest over the following $k=8$ quarters. This is conditional on candidate settings of the decision vector, which is taken as the trajectory of interest (shadow) rates over those quarters. Within-model decision analysis then delivers model-specific optimal decisions about these rates.  

\subsection{Models, Forecasts, and Model-specific Decisions\label{sec:econmodelsanddecisions}}

We consider $J=2$ models: $\cM_1$ is a three-variable monetary policy VAR involving GDP, prices, and the interest rate;  $\cM_2$ is similar to the model of \cite{frs2019}, a VAR with the same variables as $\cM_1$ plus investment/GDP ratio, stock prices, and GZ credit spread. Following \cite{frs2019} we include five lags in the VARs, and each model is identified using the sign restrictions from Table~1 of that reference. In $\cM_1$, these restrictions define supply, demand, and monetary policy shocks. In $\cM_2$, investment and financial shocks are additionally identified. We condition on a given value of the policy rate and set  monetary policy to be the driving shock. We do this by imposing restrictions on the set of structural shocks underlying the conditional forecasts. Structural shocks other than the monetary policy shock have zero means. We use the asymmetric conjugate prior of~\cite{C2022QE}, with the advantage that the marginal likelihoods for each can be easily calculated; prior hyperparameter choices are made to maximize the marginal likelihood as in this referenced paper. At each quarter, multi-step ahead predictions are based on simulations using the precision-based sampler of~\cite{chan-cond-forecast}. Details on the conditional forecast computations are summarized in 
Appendix~B. In short, this generates $p_j(\y|\x)$ with zero-mean constraints on all shocks apart from the monetary policy variable. We do not restrict the variance (i.e., \lq\lq soft" restrictions) such that we also have uncertainty around the path of $\x$. This can be thought of as conditional commitment-- we allow the possibility of $\x$ deviating from the proposed policy path with deviations informed by historical uncertainty around forecasts of the interest rate outcomes. 

It is worth noting that using models such as VARs for policymaking is sometimes questioned due to the Lucas critique. However, papers such as \cite{LEEPER2003} argue in favor of VARs as long as the policy interventions being studied are modest. We ensure BPDS will favor such modest interventions through two mechanisms. First, as noted in Section \ref{sec:DDS}, we choose only modest improvements in the target score. Second, large policy interventions will be down-weighted through  the initial probabilities reflecting the plausibility of an intervention; see details in Section~\ref{sec:informativeconditioningonx}. Thus BDPS allows for policy analysis using VARs by emphasizing decision scenarios which represent \lq\lq modest interventions", circumventing the Lucas Critique. This is an important practical contribution to the VAR literature on policy and scenario analysis.

To find optimal decisions requires choice of utility function. The form adopted here is as follows.  In the current quarter, $\x = (x_1,\ldots,x_k)'$ is the $k-$vector of interest rate values over the next $k$ quarters, and $\y = (y_1,\ldots,y_k)'$ is the corresponding $k-$vector of inflation rates. Whatever other variables are in  $\cM_j,$  interest focuses on the implied  $p_j(\y|\x,\cM_j)$ required for BPDS~\eqn{BPDSmixgeneral}. This conditional predictive is used in decision analysis with the same utility function for each model, namely $u_j(\y,\g,\x) = U(\y,\g,\x)$ given by 
\beq{DSutility} 
U(\y,\g,\x) =   - \sum_{h = \seq1k}\{\theta(y_h - y^{*})^2 + (1-\theta)(g_h - g^{*})^2 +  (x_h - x_{h-1})^2 \} 
\eeq
where $\g$ denotes GDP growth. 

This is a conventional utility function that reflects the dual mandate of inflation rate targeting, moderate growth in real-activity, and interest rate smoothing over the next $k=8$ periods. The $\y$ terms relate an inflation targeting mandate of $y^{*}=2\%$ over the longer run. The $\g$ terms reflect GDP growth with a target of $g^{*}=2.5\%$, roughly the average growth rate of GDP from 1990 to 2024. The $\x$ terms encourage relatively constrained changes in quarter-to-quarter interest rates (a \lq\lq don't rock the boat" consideration, as large swings in interest rates can/will have otherwise undesirable effects on the macroeconomy). The model-specific optimal decision vector $\x_j$ then maximizes $E_{p_j}[ U(\y,\g,\x)|\x]$ over $\x.$ Again, this analysis is repeated each quarter over time, producing rolling updates of the \lq\lq currently optimal" projections for interest rates over the coming eight quarters.  

\subsection{BPDS Model Specification}

BPDS requires specification of a relevant class of baseline pdfs $p_0(\y|\x)$, the model-specific vector score functions 
$\s_j(\y,\g,\x),$ the initial BPDS model probabilities $\pi_j(\x)$ as functions of candidate decisions $\x,$ and the target expected scores $\m_f(\x)$ at any $\x$. These are discussed in turn. Beyond customizing BPDS to the specific application, we highlight  methodological developments relevant to other applications. This includes, in particular: (a) the linkages of the $\pi_j(\x)$ to $\x$ that are relevant more generally when $\x$ is both an outcome to be forecast as well as a putative decision variable;  see Section~\ref{sec:init_prob};  (b) the relevance of dependence structure among the elements of the vector score under the initial distribution $p(\y,\cM_j)$; see Section~\ref{sec:econtargetscore}). 
Notation considers the target variable $\y$ as including both inflation and GDP 
unless specifically denoted (as in the score/utility functions).

\subsubsection{Baseline Distribution}

Completing the main BPDS p.d.f. in~\eqn{BPDSmixgeneral} requires the baseline $p_0(\y|\x).$ This is taken as a multivariate~T~distribution with 10 degrees of freedom, using the location from the initial mixture 
 $p(\y|\x)$, ignoring the baseline (i.e., with $\pi_0(\x)=0)$ and the corresponding variance of that mixture inflated by 2. This defines a relevant, tractable $\cM_0$ that can capture outcomes $\y$ that the set of VAR models are not predicting well for any $\x$ under consideration, and signal that to the decision-maker.

\subsubsection{BPDS Score Functions}

The dual mandate and interest rate smoothing from the model-specific decision analysis in Section~\ref{sec:econmodelsanddecisions}
are reflected in the BPDS score functions. We take  
\begin{align}
s_j(\y,\g,\x)= [s_{j1}(y_1, x_1), s_{j1}(g_1, x_1) \ldots, s_{jk}(y_k, x_k) s_{jk}(g_k, x_k)]'
\end{align}
with elements
\begin{align}\label{sc_bounded}
    s_{jh}(y_h, x_h) =  \exp{\{-(y_h - y^{*})^2/(2z_{y}^2)}\}  +  \exp{\{-(x_h - x_{h-1})^2/(2z_{x}^2)}\},  \quad h=\seq 1k,\\       
    s_{jh}(g_h, x_h) =  \exp{\{-(g_h - g^{*})^2/(2z_{g}^2)}\}  +  \exp{\{-(x_h - x_{h-1})^2/(2z_{x}^2)}\},  \quad h=\seq 1k,       
\end{align}
where $y^{*}= 2\%$ is the inflation target, $g^{*}= 2.5\%$ is the GDP growth target, and $z_{y}$, $z_{g}$ and $z_{x}$ are {\em score bandwidth} parameters. 
This defines a class of bounded score functions, always relevant in decision analysis and here ensuring that the entropically tilted BPDS p.d.f. of~\eqn{fj} is always integrable. 
The score bandwidths are set so that a certain deviation $d_y = (y - y^*)^2$ has a score of $\bvarepsilon$; given a choice of $\bvarepsilon$, we set $z_y = d_y/\sqrt{-2\log(\bvarepsilon)}$. Similar considerations apply to choosing $z_{x}$ and $z_{g}$. Our analyses use $\bvarepsilon = 0.4$, $d_y = 2$, $d_g = 2$, and $d_x = 1$ to ensure the score function is dispersed enough to accommodate modest changes in the Federal Funds rate while being more lenient in deviations from the inflation target. Obvious modifications could incorporate horizon $h-$specific inflation targets and differentially weight the two exponential terms, but this form suffices for our main goals in this paper. Note also that, if inflation deviations from target and interest rate changes are \lq\lq small," then $s_{jh}(y_h, x_h) $ is approximately quadratic in
$|y_h - y^{*}|$ and $|x_h - x_{h-1}|$ for all $h,$ perhaps a more familiar utility form. 

\subsubsection{Initial and Conditional Model Probabilities \label{sec:init_prob}}

For clarity in this section, we make explicit the dependency on time, so that the ingredients of the full BPDS predictive p.d.f. in 
\eqn{BPDSmixgeneral}-- with the exponential form of the calibration function of~\eqn{alphajBPDS}-- are now indexed by current time $t$; that is, 
$$
f_t(\y_t|\x_t) \propto \sum_{j=\seq 0J} \pi_{tj}(\x_t) {\textrm e}^{\btau_t(\x_t)'\s_{tj}(\y_t,\x_{tj})} p_{tj}(\y_t|\x_t,\cM_j). 
$$
Bayesian model weighting based on historical predictive performance with respect to defined forecast goals~\citep{LavineLindonWest2021avs} is the starting point for specification of the $\pi_{tj}(\x_t).$ The general form adopted is 
\beq{BPDSprobs}
\pi_{tj}(\x_t) \propto \pi_{tj} p_{tj}(\x_t|\cM_j), \qquad j=\seq 0J,
\eeq
subject to summing to 1 over $j=\seq 0J$ and with ingredients as follows. 

\subsubsection*{\em Initial model probabilities} 

Traditional analysis ~\citep[e.g.,][chapter 12]{WestHarrison1997} defines the time $t$ initial model probabilities as Bayesian updates from those at $t-1$; that is, $\pi_{tj} \propto \pi_{t-1,j} p_{tj}(\z_{t-1,j}|\cM_j)$ where the \lq\lq model marginal likelihood" term $p_{tj}(\z_{t-1,j}|\cM_j)$ is the value of the one-step ahead predictive p.d.f. under $\cM_j$ at the observed values of the last period outcomes $\z_{t-1,j}$ under that model.  In our setting, this includes time $t-1$ outcomes of inflation $(y)$, interest rate $(x)$, and other economic indicators in $\cM_j$. In general these can differ across models, but in consideration for the initial weights we restrict to variables common across models. 

BPDS allows the decision-maker freedom to make alternative choices of the $\pi_{tj},$ and the goal and decision focus recommend modification of the standard BMA choice. BMA, after all, only reweights models based on one-step ahead predictive accuracy. Hence, we adopt two modifications based on recent literature consonant with the goal foci. 

First, we use simple power discounting of historically accrued support across models, in which the time $t-1$ to time $t$ evolution is reflected in
$\pi_{tj}\propto \pi_{t-1,j}^\gamma p_{tj}(\z_{t-1,j}|\cM_j)$, where $\gamma$ is a discount factor in $(0,1]$, closer to 1 for most applications. This acts to discount historically accrued support for model $j$ at a per-time unit discount rate $\gamma$ prior to updating by the time $t-1$ information. Going back at least to~\cite{JQSmith1979} and then, in a formal dynamic model uncertainty context,~\citet[][chapter 12, p. 445]{WestHarrison1989}, power-discounting has been shown to be of value in empirical studies in implicitly allowing for time variation in the predictive relevance of different models~\citep[e.g.,][]{Raftery10,Koop2013,ZhaoXieWest2016ASMBI}. Our case study uses $\gamma = 0.95$. 
  
Second, initial model probabilities are modified based on the recent relative performance of models with respect to the defined goals. This is the premise underlying the specific variants of BPS in the setting of adaptive variable selection, or BPS-AVS, in ~\cite{LavineLindonWest2021avs}, and related developments in~\cite{LoaizaMayaJOE2021}, for example. This leads to the immediate BPDS extension of these prior approaches with
$$\pi_{tj}\propto \pi_{t-1,j}^\gamma p_{tj}(\z_{t-1,j}|\cM_j) \e^{\btau_{t-1}(\x_{t-1})' \s_{t-1,j}(\y_{t-1}, \x_{t-1,j})}.$$
Here the discounted past probabilities are updated with AVS-style weights using the realized BPDS calibration function with 
relative model scores based on the actual decision outcomes at the last time period. As a result, models are initially and naturally reweighted based on both predictive and decision outcome performance at the last time period. 
 
The result at this stage is that models achieving \lq\lq good" recent trajectories of  interest rate smoothness, as well as relatively accurate forecasting performance of realized inflation outcomes, will be rewarded with higher initial BPDS model probabilities looking forward.    Of course, the specification can cut back to define special cases including BPS-AVS (with $\btau_{t-1}(\x_{t-1})=\bzero)$, and within that to traditional BMA (with $\gamma=1$) for comparisons. 

 Finally, the inclusion in BPDS of the baseline model and its forecast densities leads to a modification of these initial model probabilities to provide a non-zero value $\pi_{t0}$ for the baseline. We choose a fixed probability-- in our analysis $\pi_{t0} =0.1$ at each time point $t$-- and simply renormalize the $\pi_{tj}$ over $j=\seq 1J$ accordingly.

\subsubsection*{\em Informative conditioning on $\x_t$\label{sec:informativeconditioningonx}}  
 
As noted earlier, in our setting the future values of decision variables are also considered outcomes predicted under the models.  The models each 
forecast the future evolution of interest rates as part of the complex, dynamic macroeconomic system, whereas for decisions we must condition on $\x_t$. This is reflected in the conditional (on $\x_t$) distributions $p_{tj}(\y_t|\x_t,\cM_j)$ in BPDS where $\x_t$ is treated as known. 
The theoretical implication for the BPDS model probabilities is the term
  $p_{tj}(\x_t|\cM_j) $ in~\eqn{BPDSprobs}-- this is the value of the current marginal predictive p.d.f. of the vector $\x_t$ under $\cM_j.$
Assuming the prior (to time $t$) probabilities $\pi_{tj}$ are specified, this form arises directly via Bayes' theorem. The act of conditioning on $\x_t$ is informative, and the implied update is, simply by Bayes' theorem, that in~\eqn{BPDSprobs}.
Critically, this implies that candidate decision values that are not well-supported under the joint distribution of a model are down-weighted. Conversely, at any candidate decision vector $\x_t,$ models that are more predictively supportive of the decision $\x_t$ will be relatively rewarded with higher values of resulting $\pi_{tj}(\x_t)$. 

This is keeping with \cite{LEEPER2003} who argue that reduced form models are appropriate to use for policy analysis if the policy does not deviate too much from past policy changes. If the decision is very unusual agents may think they are in a new policy regime and adjust their behavior rendering a reduced form model less useful (i.e. the Lucas Critique). The technique above addresses these concerns by explicitly increasing the weight on models where the policy decision represents only a \lq\lq modest intervention'. Further, if the decision is unsupported by the model set then the baseline model, being dispersed with fat tails, will receive more weight. In this way BPDS up-weights models least susceptible to the Lucas Critique, and in periods when there is more uncertainty the weights increase on the baseline model.
  
In other applications of BPDS, the decision variables may be exogenous; that is, control variables that are to be chosen by the decision-maker but that are not forecast jointly with $\y_t$ in the set of models. In such cases, it will be common to assume that the external choice of $\x_t$ is not informative, and then~\eqn{BPDSprobs} results in decision-independent BPDS probabilities $\pi_{tj}(\x_t) =\pi_{tj} $ based only on historical data and information.

\subsubsection{BPDS Target Scores\label{sec:econtargetscore}}
The BPDS  expected score $\m_f(\x) = E_f[\s(\y, \x)]$  represents a target for improvement over the initial expected score $\m_p(\x)=E_p[\s(\y, \x)]$. In the multi-objective case, the resulting $\btau(\x)$ that defines $f(\y|\x)$ to satisfy this target expectation is sensitive to both the relative scales and dependence of elements of $\s(\y, \x)$ under the initial mixture $\y \sim p(\y|\x)$ at any candidate decision $\x.$ As functions of $\y$, the elements of the random score vector $\s(\y, \x)$ can be strongly correlated, leading to challenges in specifying relevant targets. This can also complicate the calculation of the implied BPDS tilting vector $\btau(\x)$ (i.e., the vector that is needed to satisfy $\m_f(\x) = E_f[\s(\y, \x)]$ under the BPDS density of~\eqn{BPDSmixgeneral}).
We address this by explicitly recognizing score dependencies and defining an approach that explicitly incorporates dependence. 

Some theoretical intuition is gained by considered cases of \lq\lq small perturbations" in which $\m_f(\x) - \m_p(\x)$ has small elements. In this setting, entropic tilting theory in~\cite{TallmanWestET2022} yields the second-order approximation 
$\btau(\x) \approx \V_p(\x)^{-1} (\m_f(\x)-\m_p(\x))$ where $\V_p(\x)$ is the variance matrix of $\s(\y,\x)$ under 
the initial mixture $p(\y|\x).$ 
This shows that the implied tilting vector will be very sensitive to the initial score scales and dependencies as reflected in $\V_p(\x),$ and suggests a prime focus on a {\em standardized score} scale; that is, define $\C_p(\x)$ as the scaled eigenvector matrix such that $\V_p(\x)=\C_p(\x)\C_p(\x)'$ and set the target score using
$\m_f(\x) = \m_p(\x) + \C_p(\x)\bepsilon(\x)$ for a specified {\em standardized expected score vector} $\bepsilon(\x)$. The usual convention is taken in which the eigenvector columns of $\C_p(\x)$ are ordered according to decreasingly values of the corresponding eigenvalues, so that the first column is \lq\lq dominant," 
and so forth. This provides insights into how to practically define target scores related to the absolute standardized scale. As examples of the two extremes, taking 
$\bepsilon(\x) = \epsilon(\x)\bone$ for some scalar $\epsilon(\x)$ represents targets deviating from the initial expected score in equal amounts of $\epsilon(\x)$ along each of the standardized eigen dimensions. At the other extreme, and most relevant when there are strong score dependencies, taking $\bepsilon(\x)= (\epsilon(\x),0,\ldots,0)'$ defines the resulting target $\m_f(\x)$ based on the major, dominant eigen dimension alone.  The latter is a starting point in general and is taken to define our BPDS case study that follows. In that setting, we choose $\epsilon(\x)$ such that $\min \{(\m_f(\x) / \m_p(\x))\} = 0.75$ to define the maximum expected improved score in any dimension.\footnote{Additionally, due to the arbitrariness of the signs of eigenvectors, we apply a $\pm 1$ multiplier to the first column of $\C_p(\x)$ so that the sum of the elements are positive, ensuring the target score improves upon $\m_p(\x)$.} It is obviously straightforward to extend this methodology to define target scores impacted by higher eigen dimensions, though that is left for future applications. 
 
\subsection{BPDS Implementation and Optimal Decisions \label{sec:ComputationEtc}}

The final step couples the decision-maker's utility with the BPDS predictive eqns.~(\ref{eq:BPDSmixgeneral},\ref{eq:BPDSmixgeneralfform}) to define the optimal decisions from the model synthesis. The decision-maker can adopt any utility function, but an initial neutral analysis will be based on using the same form as usual in the model-specific decisions: the function $U(\y,\x)$ of~\eqn{DSutility}. This is used here, computing $\x$ to maximize the implied expected utility function $\bar U(\x).$ We compare decisions recommended by BPDS to those from each of the models and to the traditional BMA-based analysis.  BMA uses model weights proportional to the marginal likelihoods of the data that are common across models (including inflation, interest rate, and GDP).
As earlier discussed in Section~\ref{sec:BPDS}, BMA arises as a special case of the BPDS analysis.

The computation of BPDS involves two key components. First, the overall optimization over $\x$  explores potential BPDS decisions and finds the optimizing vector $\x$. This requires an \lq\lq outer loop" numerical optimization to explore $\x$ space. This is done using the MATLAB implementation of particle swarm optimization. We use this due to occasional multi-modality in $\bar U(\x)$ and because the algorithm supports parallelization which significantly speeds up optimization. Model specific optimizations are done using a trust region method, namely, Powell's Derivative Free Optimization Solvers (PDFO--~\citealp{pdfo}).  Second, within each evaluation of a potential BPDS decision, it is necessary to compute the tilting vector $\btau(\x)$ given the constraint $E_f[\s(\y,\x)]=\m_f(\x)$ for a target expected score $\m_f(\x).$ The theoretical basis of this is an implicit equation that is solved via standard, generic numerical optimization methods. Relevant details follow~\citet[][section 4.4]{TallmanWest2023} and are summarized in our Appendix~A.


BPDS forecast distributions are evaluated using importance sampling. At any given $\x,$ the BPDS predictive distribution 
in~\eqn{BPDSmixgeneral} is simulated by sampling from the $p_j(\y|\x,\cM_j)$ in proportions defined by the BPDS probabilities $\pi_j(\x)$. Importance sampling weights are then proportional to the realized values of $\alpha_j(\y|\x).$ 
This provides for efficient computation as well as access to traditional methods and metrics-- such as the importance of sampling effective sample sizes (ESS--~\citealp[e.g.,][in related contexts]{GruberWest2016,GruberWest2017ECOSTA})-- to monitor and evaluate the quality of the resulting Monte Carlo approximations compared to resulting predictive expectations. 
Note that this evaluation can deliver such metrics to assess \lq\lq concordance" between the initial densities $p_j(\y|\x,\cM_j)$ and their corresponding BPDS-tilted versions 
$f_j(\y|\x,\cM_j)$ in~\eqn{BPDSmixgeneralfform}, as well as that of the initial mixture $p(\y|\x)$ and the resulting $f(\y|\x).$ More aggressive BPDS target scores will generally lead to lower concordance, and choices can be partly guided by such empirical evaluations.

\section{Case Study} 

\subsection{Overview}
Analyses proceed sequentially on an expanding window of data beginning in 1992Q2. We compare decisions recommended by BPDS to those under BMA, discuss the individual models and how they are combined by BPDS and BMA, and highlight some operational BPDS details underlying insights into the resulting decision outcomes. 

\subsection{Optimal Decisions \label{sec:optdecs}}
\begin{figure}[b!]
    \centering
    \includegraphics[width=0.9\textwidth,keepaspectratio]{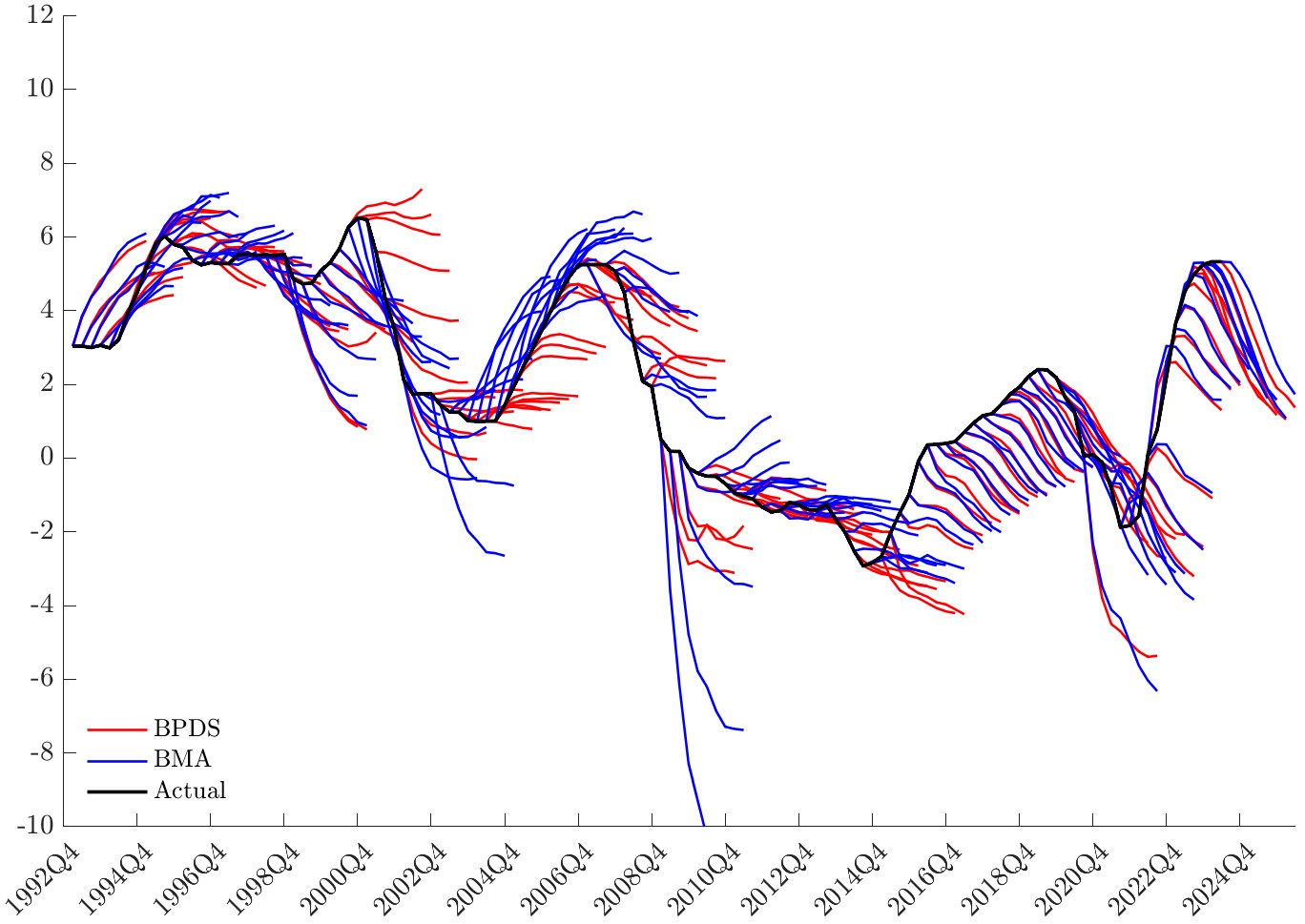}
    \caption{Recursively calculated policy decisions   
    \label{fig:hist_decisions}}
\end{figure}

Figure~\ref{fig:hist_decisions} shows the actual policy rate each quarter along with the $1{-}8$-quarters-ahead policy recommendations that would have been made by BPDS and BMA. In using the shadow rate, the zero lower bound is not in effect and negative values for the policy rate are possible. Recommendations for negative values for the policy rate are not to be taken literally as advising cuts to a negative Federal Funds rate, but rather as a suggestion to undertake other forms of monetary easing that would be expected to proxy such cuts. 

Since 2014, the optimal policy paths recommended by the two approaches are generally similar, though there are notable differences prior to that time. Some specific periods of interest are now highlighted. 

\subsubsection*{\em 2014 to the present} During this period, BMA and BPDS provide similar recommendations that are often quite different from the actual policy rate. For almost all of these times, the policy recommendations are to cut interest rates, whereas (apart from 2019--2021) the actual policy rate increased. Some differences do arise between BMA and BPDS. For example, during the post-COVID inflation period, BPDS recommends a higher rate path. BPDS is closer to the decision actually made by the Federal Reserve, although according to BPDS, interest rates should decrease throughout 2023 and 2024.  


\subsubsection*{\em The financial crisis and subsequent recession}
It is during this period that the differences between BPDS and BMA are most acute. The actual policy rate fell slowly during this period. BPDS recommends rate cuts as well, initially at a more rapid rate than what actually occurred, but as of 2010, its recommendations are similar to the ones the policymakers actually made. In contrast, BMA recommends huge cuts to the policy rate right at the start of the financial crisis, but subsequently consistently argues for rate increases.

\subsubsection*{\em The first years of the 21st century}
From 2003 through to the beginning of the financial crisis, the actual policy rate was gradually increasing. In this period, BMA consistently recommends rapid rate increases. In contrast, BPDS recommendations are generally similar to what actually transpired, apart from at the beginning of this period where the advice is to raise the policy rate more slowly than what actually occurred.

\subsubsection*{\em The 1990s}  During this period, the pattern is more mixed. Optimal policy recommendations generated under each of BMA and BPDS often differ from actual decisions, with no consistent pattern; at times the recommended rates are higher than the actual policy rate, and at other times they are lower.

A general pattern, one that occurs throughout the sample period, is that BMA and BPDS typically recommend larger changes in policy rates than were actually implemented by policymakers. Part of this is presumably due to differences between the policymakers' utility function and those used in our analyses. Other possible explanations are that policymakers can affect expectations through their communications, which is a channel not captured in the model, or their models use a much steeper Phillips Curve. Also, we focus only on inflation up to two years ahead, without considering the possibility of an over- or under-shoot of inflation after eight quarters. In contrast, policymakers would generally aim for inflation to be sustainably at target over the longer-term.

\subsection{Trajectories of BPDS and BMA Predictive Densities}

\begin{figure}[p!]
    \centering
    \includegraphics[width=0.9\textwidth,keepaspectratio]{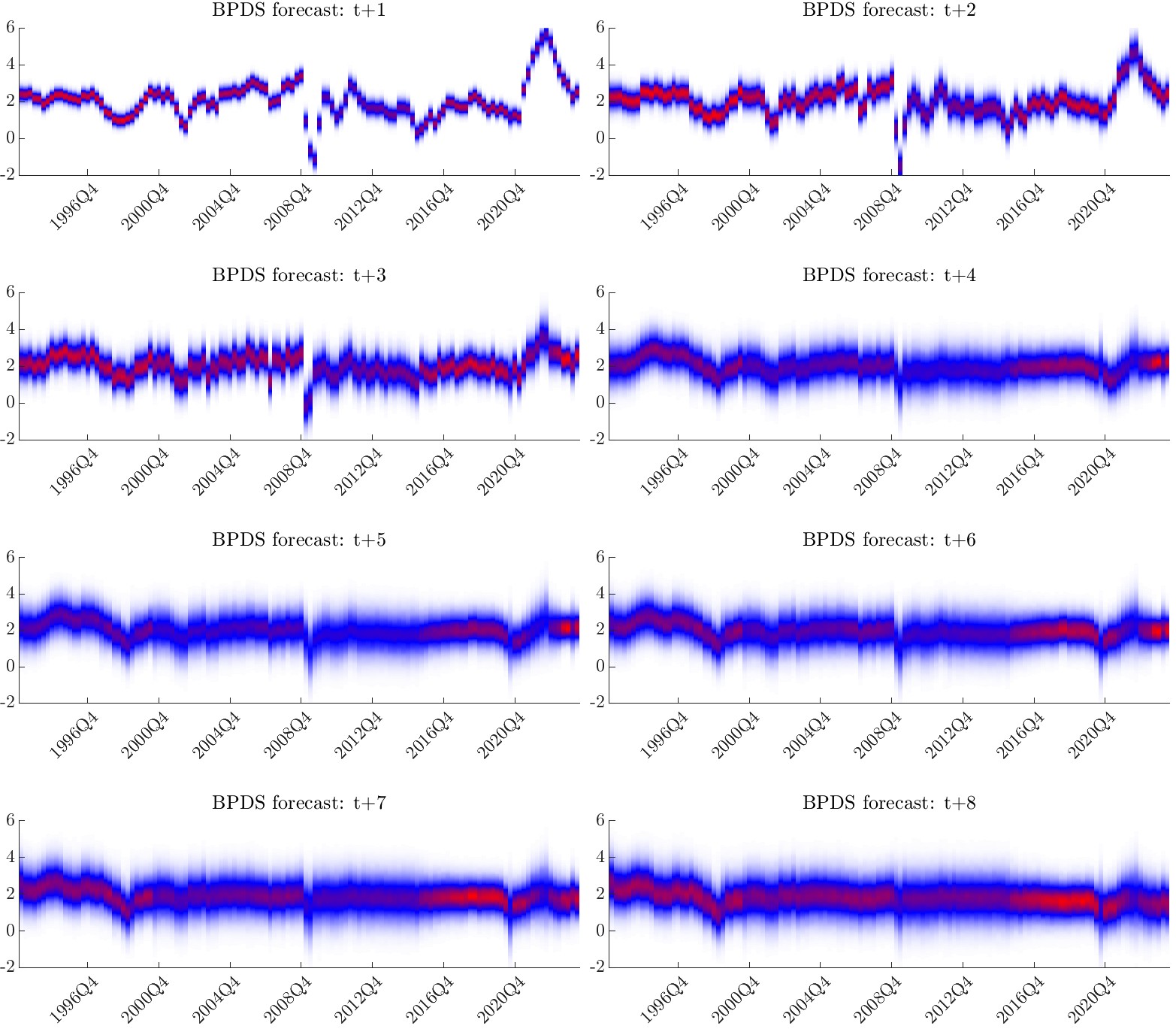}
    \caption{BPDS forecast densities of inflation. The frames represent 1--8 quarter ahead forecasts, reading along the rows from top-left to bottom-right. The colours represent probabilities, with the blue shading showing lower probability, and red showing higher probability.  \label{fig:BPDS_density}}
\end{figure}

\begin{figure}[p!]
    \centering
    \includegraphics[width=0.9\textwidth,keepaspectratio]{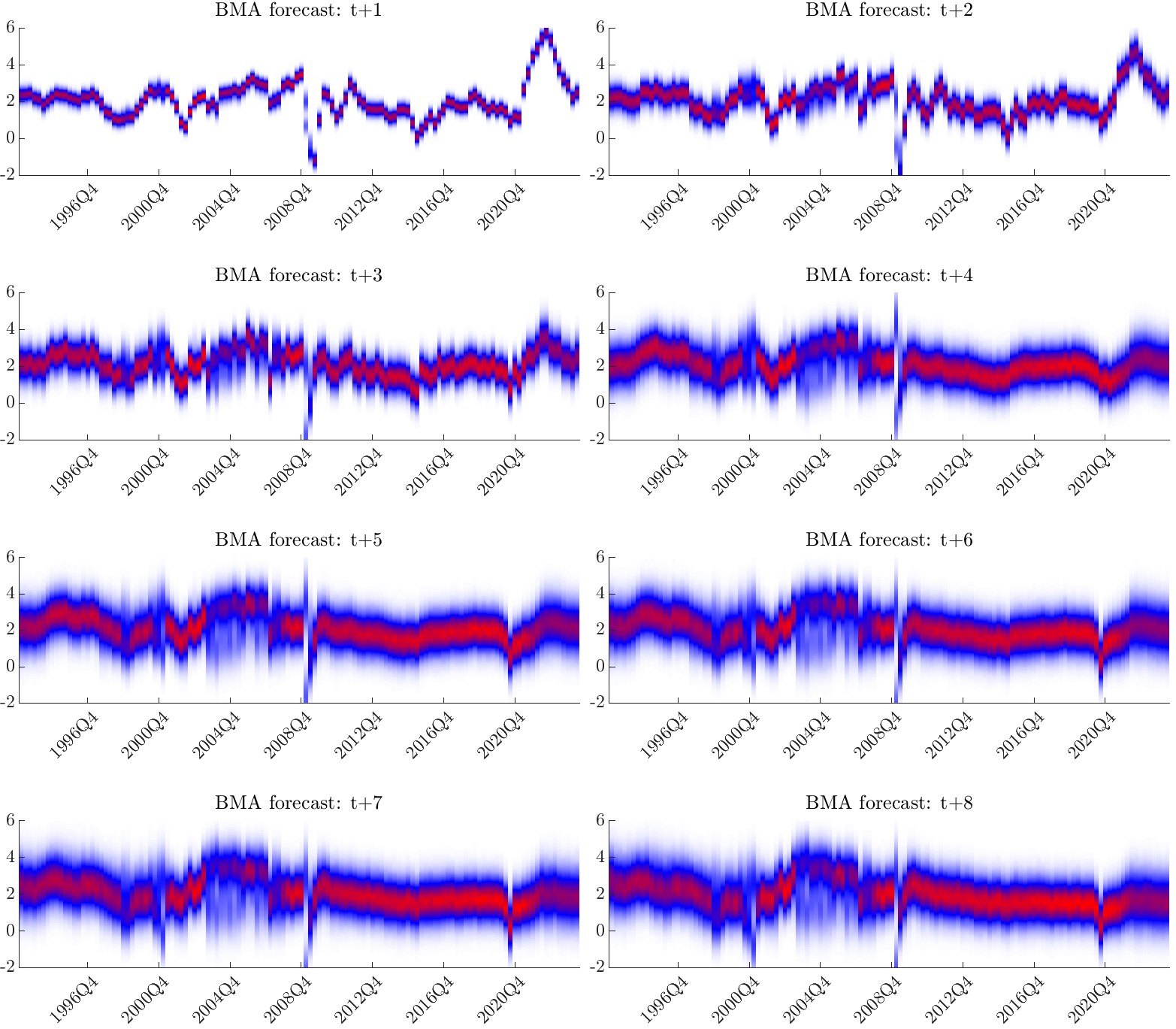}
        \caption{BMA forecast densities of inflation. The frames represent 1--8 quarter ahead forecasts, reading along the rows from top-left to bottom-right. The colours represent probabilities, with the blue shading showing lower probability, and red showing higher probability. \label{fig:BMA_density}}
\end{figure}

\begin{figure}[p!]
    \centering
    \includegraphics[width=0.9\textwidth,keepaspectratio]{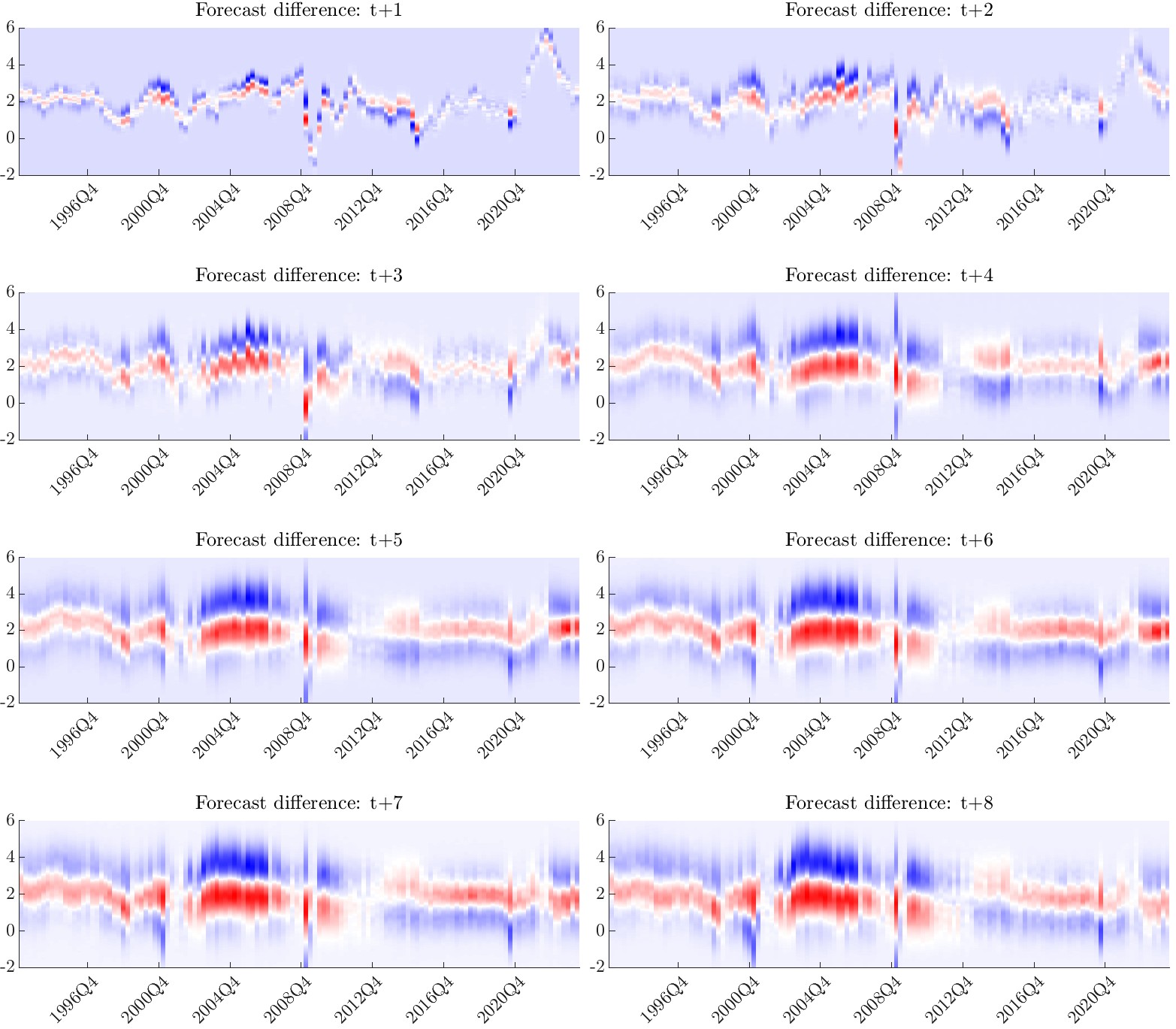}
        \caption{Difference between BPDS and BMA forecast densities of inflation. Red-shaded regions have higher probability under BPDS than under BMA, with blue shading indicating the reverse, and white shading showing equal probability. \label{fig:density_diff}}
\end{figure}

Figures~\ref{fig:BPDS_density}--\ref{fig:density_diff} shed light on these patterns. These images represent the time trajectories of predictive densities of inflation at each relevant horizon, using  BPDS (Fig.~\ref{fig:BPDS_density}) and BMA~(Fig.~\ref{fig:BMA_density}), as well as their differences~(Fig.~\ref{fig:density_diff}). These indicate that the BPDS mixtures are less dispersed than the BMA mixture for much of the sample period; that is, BMA predictive distributions are relatively more heavy-tailed, especially at longer horizons. This is partly due to the BPDS score function emphasizing that the policymaker wants to avoid extreme inflation outcomes, and also accounts for why BMA often tells policymakers to make larger changes to the policy rate than BPDS, as discussed in Section~\ref{sec:optdecs}. The differences between BPDS and BMA become larger at longer forecast horizons. Medium- and long-term macroeconomic forecasting is difficult, which leads to standard methods such as BMA producing fairly dispersed predictive densities at longer horizons. BPDS, on the other hand, is reducing this effect, which dampens the BPDS optimal decisions and reduces predictive uncertainty relative to BMA. Then, differences between BPDS and BMA forecast densities are reduced after the financial crisis, which helps account for why their policy recommendations are similar in the last decade of the sample.

\subsection{Model Probabilities}

\begin{figure}[b!]
    \centering
    \subfigure[]{\includegraphics[width=0.45\textwidth,keepaspectratio]{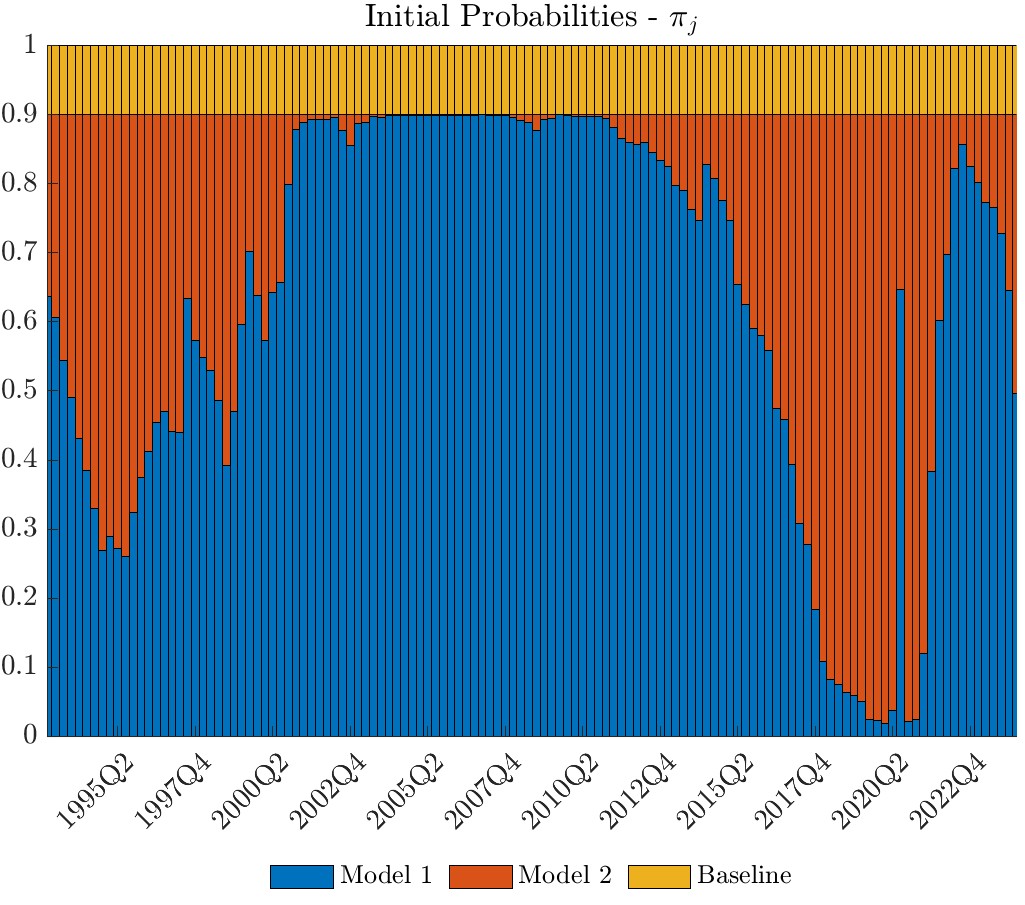}}
    \subfigure[]{\includegraphics[width=0.45\textwidth,keepaspectratio]{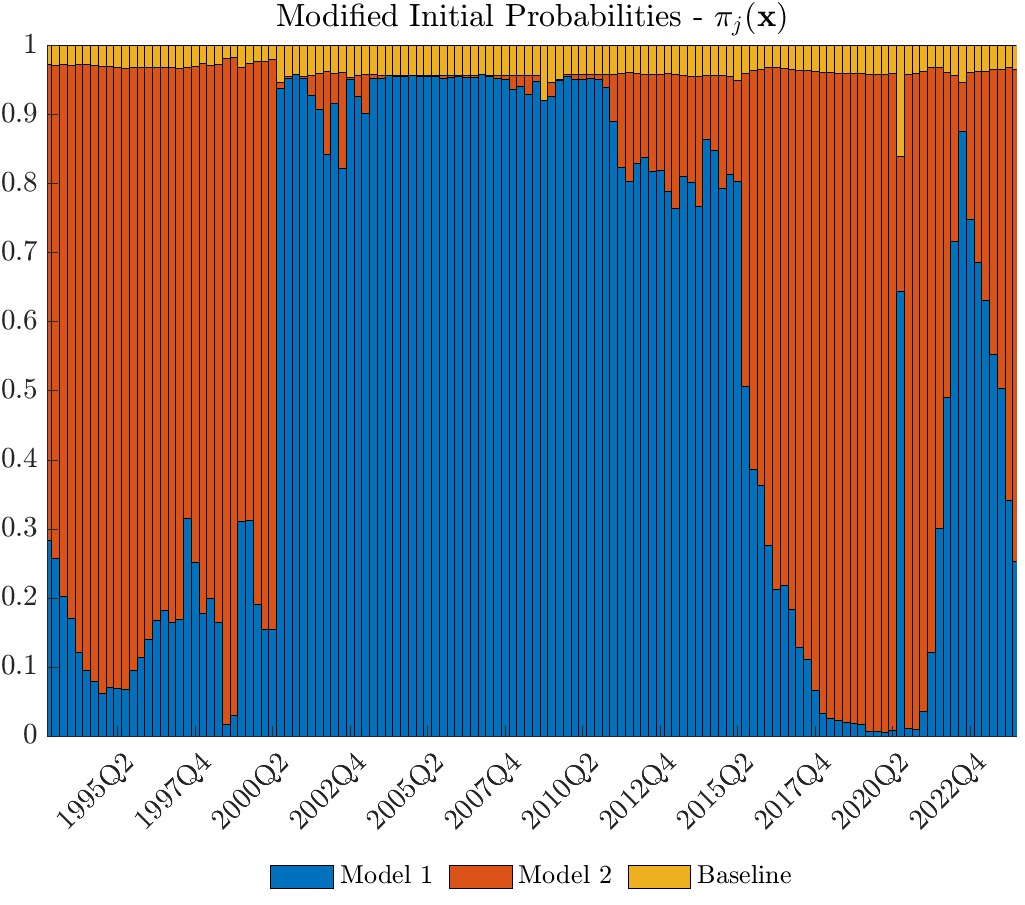}}
    \subfigure[]{\includegraphics[width=0.45\textwidth,keepaspectratio]{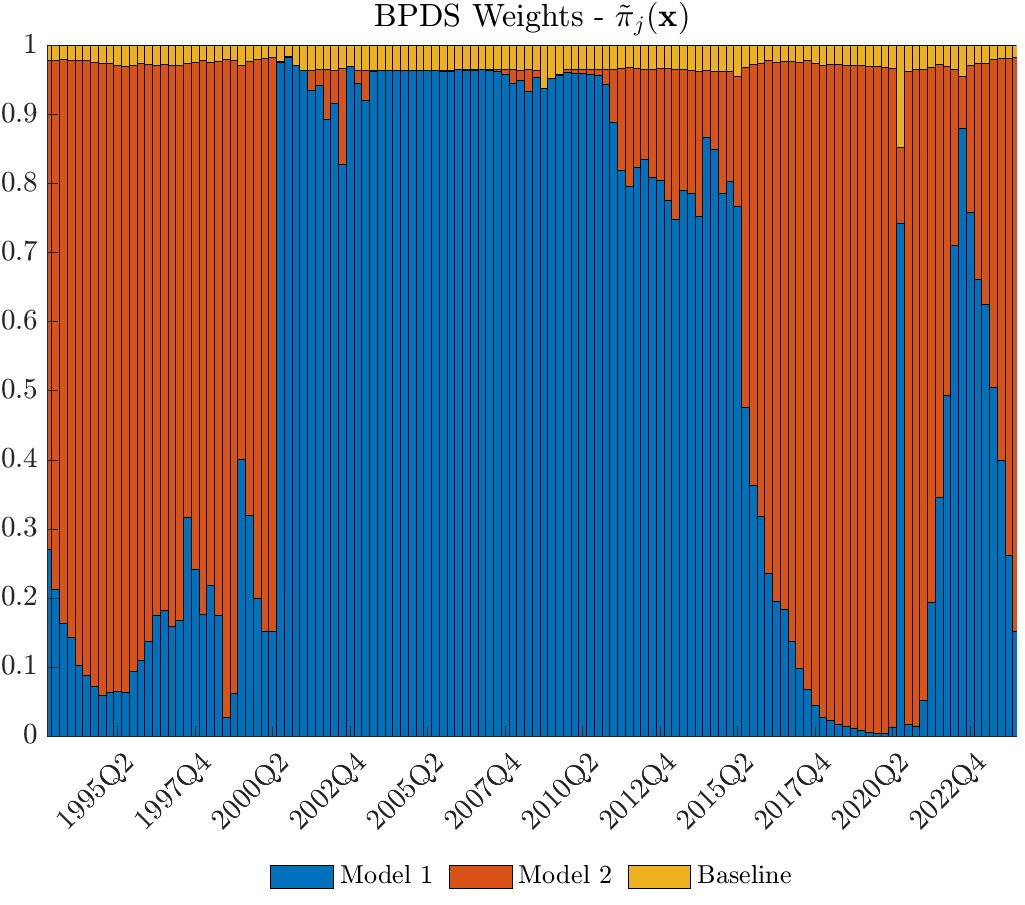}}
    \subfigure[]{\includegraphics[width=0.45\textwidth,keepaspectratio]{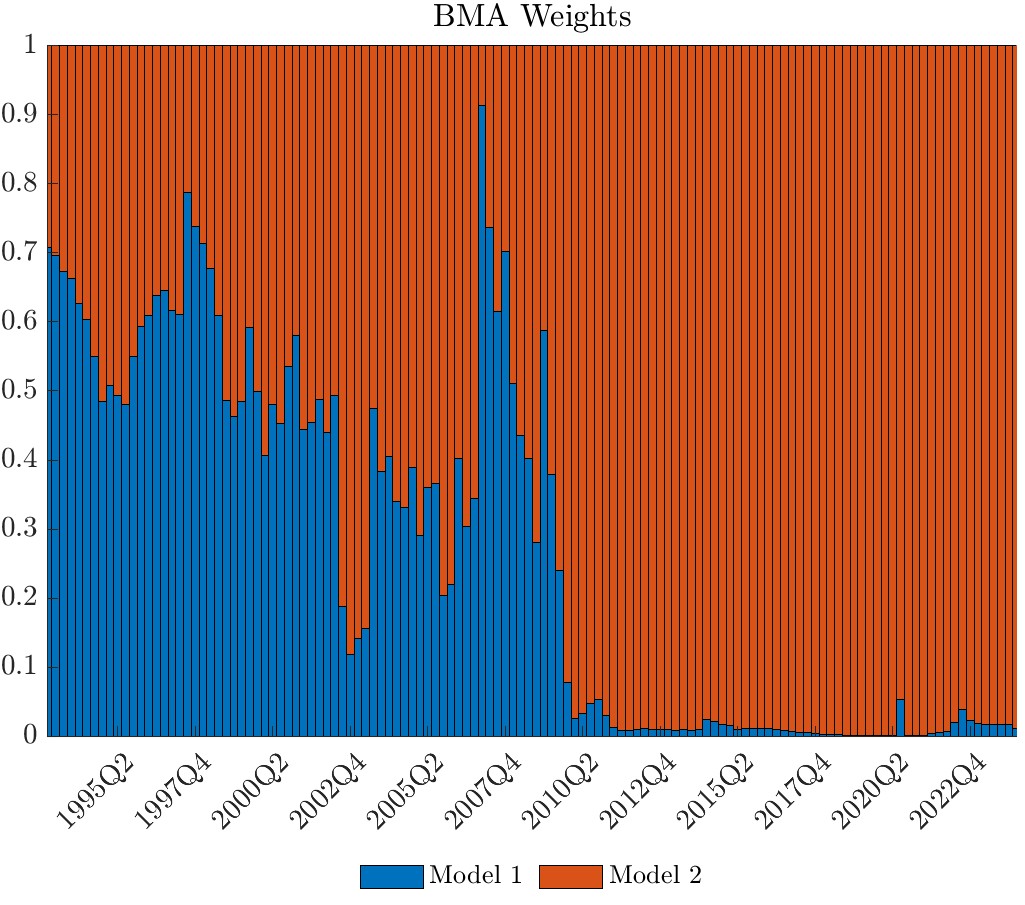}}
        \caption{Time trajectories of model probabilities. (a) Prior BPDS probabilities $\pi_{tj}$ based on discounted AVS with a fixed baseline $\pi_{t0}=0.1;$ (b) BPDS decision-dependent initial probabilities $\pi_{tj}(\x_t); $ (c) Implied BPDS weights $\tilde\pi_{tj}(\x_t);$ (d) BMA probabilities. 
     \label{fig:model_prob}}
\end{figure}

Figure~\ref{fig:model_prob} shows trajectories of model probabilities under BPDS and BMA. At each $t$, these are the discounted AVS prior model probabilities $\pi_{tj}$,  implied initial decision-dependent probabilities $\pi_{tj}(\x_t)$ evaluated at the BPDS-optimal decision $\x_t$,  resulting 
BPDS probabilities $\tilde\pi_{tj}(\x_t)$ of~\eqn{fj}, and  standard BMA probabilities.  

Under traditional BMA, the two model probabilities are appreciable until the financial crisis. After the start of the crisis, the less parsimonious $\cM_2$, which includes additional financial variables, receives virtually all the weight. In contrast, BPDS weights vary more over time, allocating most of the weight to the parsimonious $\cM_1$ for much of the period (i.e., 1997 through 2017), though $\cM_2$ plays more of a role at both the beginning and end of the sample period. That BPDS generally favours the more parsimonious $\cM_1$, with less dispersed forecast distributions,  partially accounts for why BPDS often dampens extreme recommendations made when using BMA.

BPDS probabilities on the over-dispersed $\cM_0$ are generally small, though with notable increases at the start of the COVID-19 pandemic. In such extreme times, when neither $\cM_1$ nor $\cM_2$ forecasts well, the increased probability on the fall-back $\cM_0$-- though small-- provides an indicator of this. 

The BPDS prior model probabilities $\pi_{tj}$ based on discounted AVS differ noticeably from BMA probabilities (except at the very start of the time period). A big impact then arises from the conditioning on information generated in the decision space to map these $\pi_{tj}$ to the decision-dependent weights $\pi_{tj}(\x_t)$ at the BPDS optimal decisions $\x_t$ at each time.   Recall that this mapping theoretically properly takes into account the likelihood of future, as yet unobserved, interest rate outcomes; this relevant information is not accounted for in the prior weights $\pi_{tj}$ and is, of course, absent under BMA. The subsequent map from initial probabilities $\pi_{tj}(\x_t)$ to the BPDS weights $\tilde\pi_{tj}(\x_t)$ is wholly based on the impact of the entropic tilting towards \lq\lq more favourable" decisions in expectation. We see that the impact is rather small over time, and this is to be expected: the BPDS analysis uses \lq\lq small" perturbations of the initial mixture based on target expected scores that are only modest increases over those under the initial mixture. We expect to see slight tilting towards models that are expected to do well, but not large changes relative to the initial probabilities.

\subsection{Additional Insights from BPDS Results}

\begin{figure}[b!]
    \centering
    \subfigure[]{\includegraphics[width=0.45\textwidth,keepaspectratio]{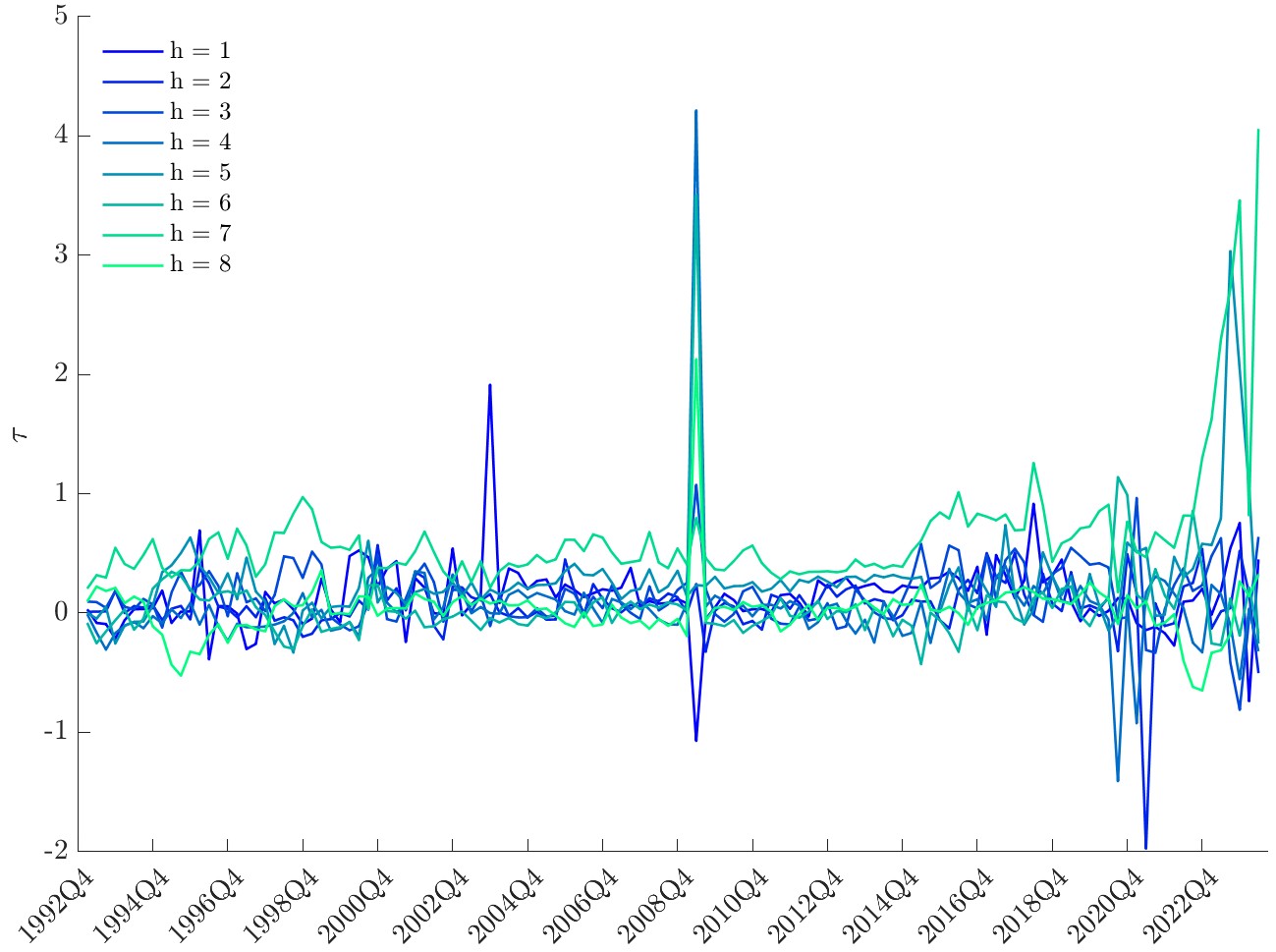}}
    \subfigure[]{\includegraphics[width=0.45\textwidth,keepaspectratio]{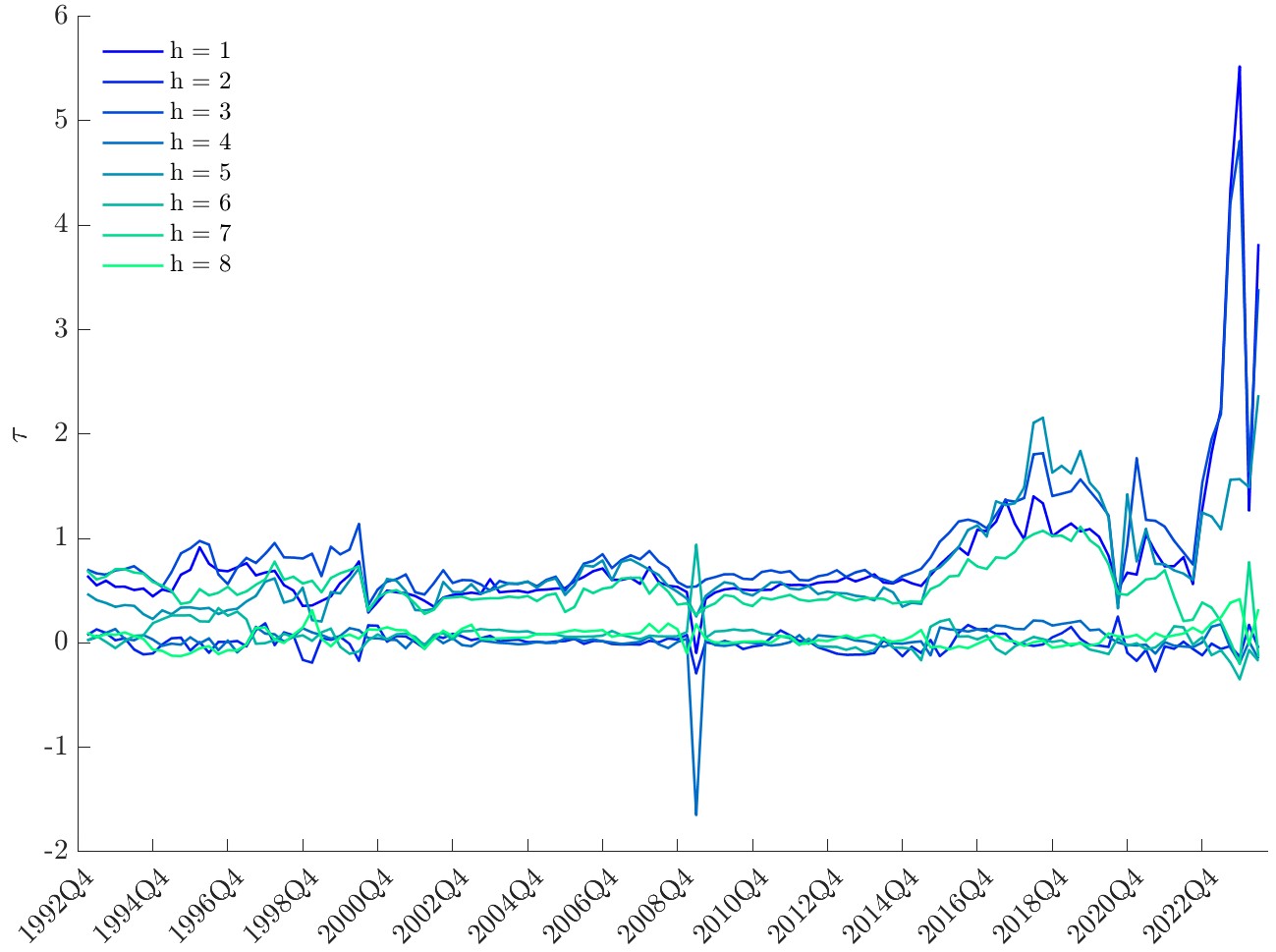}}
        \caption{Trajectories of the sixteen elements of the evaluated BPDS tilting vector $\btau_t(\x_t)$ at the optimized $\x_t$ at each quarter. (a) $\btau_t(\x_t)$ associated with inflation (b) $\btau_t(\x_t)$ associated with GDP. 
     \label{fig:tau}}

\bigskip\bigskip
    \centering
    \includegraphics[width=0.6\textwidth,keepaspectratio]{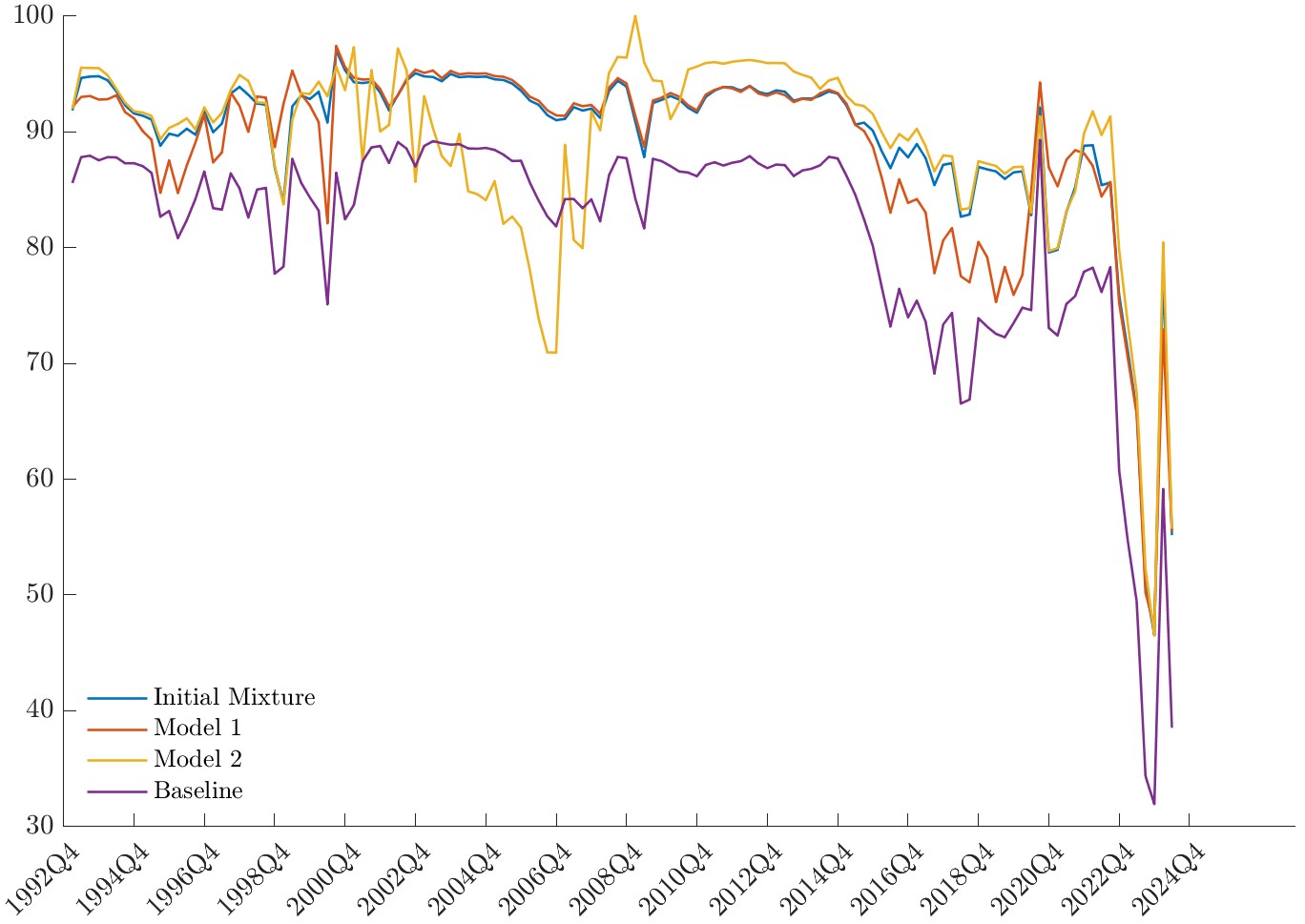}
    \caption{Trajectories of the effective sample size (ESS) metrics for individual models and for the BPDS initial mixture. \label{fig:ESS}}
\end{figure}

Time trajectories of the tilting vectors $\btau_t(\x_t)$ evaluated at the optimal decisions $\x_t$ are shown in Figure~\ref{fig:tau}. The values generally tend to increase with horizon $h$, thus attaching more weight to longer forecasting horizons. This is partly to be expected due to the higher uncertainties at longer forecast horizons.

Figure~\ref{fig:ESS} plots trajectories of several effective sample size (ESS) measures arising from the importance sampling to simulate BPDS predictive distributions, as discussed in Section~\ref{sec:ComputationEtc}. This provides a read-out of the extent of tilting the initial mixture $p(\y|\x)$ to the BPDS mixture $f(\y|\x)$, as well as that for tilting each of the individual model pdfs from $p_j(\y|\x,\cM_j)$ to $f_j(\y|\x,\cM_j)$ (again, time-indexed and updated throughout the time series). Until the COVID recession, the ESS of the initial mixture is stable between 90--95\%, suggesting only a small amount of tilting, as desired. The COVID recession is a period of rapid change, as expected, as we see large changes in the initial weights and larger values of $\btau$ required to achieve the desired target. The low value of ESS indicates that at that point, the target expected scores are unrealistic given the then-current state of the economy. However, the resulting decisions during this time period appear to be rather sensible. This means we do not need to be too concerned about the low ESS, which can, in any case, be redressed by simply increasing the overall Monte Carlo sample size accordingly. The ESS values of individual models are generally lower than that of the overall mixture and somewhat more volatile. One nice point is that, even when one of the models seems to suffer a low ESS, the BPDS mixture ESS is generally maintained at higher values. This indicates that BPDS 
is able to strike a balance in weighting expected versus historical performance of models on both predictive and decision outcomes.

\begin{figure}[tb!]
    \centering
        \includegraphics[width=0.6\textwidth,keepaspectratio]{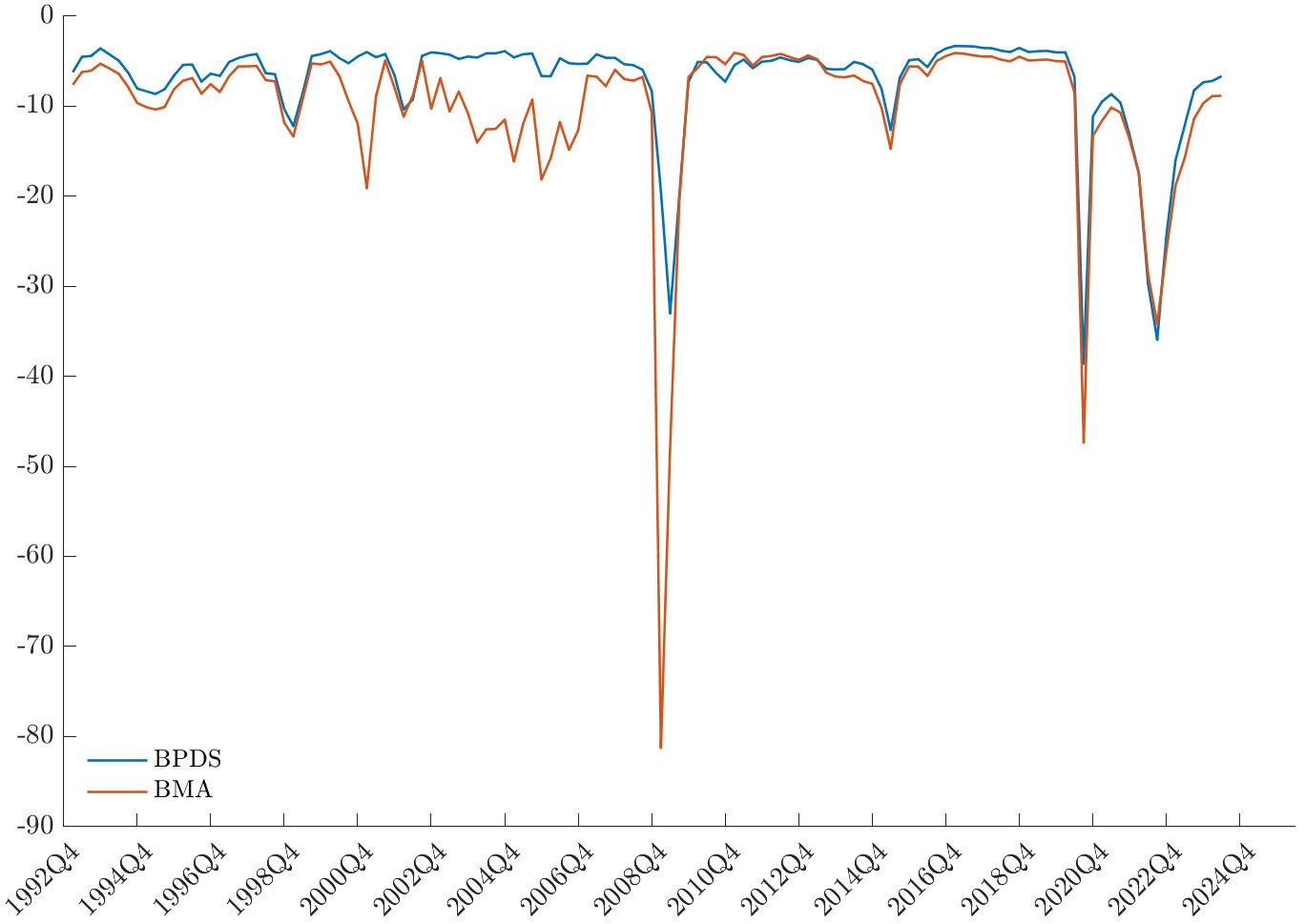}
    \caption{Trajectories of expected utilities comparing BPDS with BMA.
   \label{fig:util}}
\end{figure}

Finally, Figure~\ref{fig:util} compares the realized trajectories of expected utilities under BPDS and BMA. Each uses the same utility function to define the final optimal policy path decision, so these are directly comparable, and the comparison is relevant in terms of the setting of forward, sequential decisions where a change to much lower values at any time point should signal concern to the decision-maker. BPDS is designed to target an expected utility higher than that of the initial mixture, but whether it achieves a higher expected utility than BMA-- which has different initial probabilities and lacks outcome-dependent weighting-- is a question for empirical study. In this example, as illustrated in the figure, BPDS utility does exceed that of BMA in virtually every period. After the financial crisis, the two are similar, consistent with the earlier finding that they typically produce similar decisions during this time period. However, before the financial crisis, there are several periods during which the BPDS expected utilities are substantially higher than those of BMA. These correspond to times where we see more differences between optimal policy path recommendations. Within these times there are some periods of greater concordance between BPDS and actual policy decisions, as well as more constrained (i.e., less extreme) recommended decisions under BPDS relative to BMA. 
 
\section{Summary Comments}

BPDS is the formal, foundational Bayesian framework that extends traditional Bayesian model uncertainty analysis to address explicit use of model-specific decision outcomes as well as purely predictive performance in model comparison and combination. 
This paper has adapted the BPDS foundations to define implied methodology in formulating macro-economic decision-making when faced with multiple objectives and multiple outcomes of interest in the monetary policy setting. 

Earlier applications of BPDS  focused on financial forecasting and portfolio decisions~\citep{TallmanWest2023,TallmanWest2024}. In this setting, forecasting models do not (generally) depend on the decisions of interest, while utility functions may and often do depend on the models and their predictions. In contrast, the setting of monetary policy analysis is one in which the dependence of models and their forecasts on the decision variables (policy instruments) is simply fundamental.  It is also a setting in which the decision variables are treated simultaneously as outcomes. The future paths of central bank interest rates, for example, are modelled as time series outcomes along with other economic and financial indicators in VAR models. This then leads to conditioning on decision variables to define predictions of other indicators, with consequent implications for relative model weights in the model uncertainty setting. This latter point is critical as it then leads to relatively up- or down-weighting a model based on how well-supported a particular candidate decision is under its predictions; to our knowledge, this is the first time this central question has been formally, statistically addressed. These central features of predictive decision-making in monetary policy contexts are addressed with extensions and customization of the existing theory of BPDS. 

The BPDS perspective-- of integrating historical and expected decision outcomes with focused aspects of statistical predictive performance into relative 
model weightings-- is new to the policy arena. We argue for this perspective since policymakers are primarily interested in using sets of models for the eventual policy decisions. Pure forecasting exercises-- and evaluation and combinations of models for prediction {\em per se}-- are, of course, of parallel interest and importance. We emphasize that BPDS also involves addressing predictive performance on specific, defined outcomes of interest.  But most importantly, by putting the spotlight on decision-making, we gain additional insights into policy-making that are not possible in exercises that focus solely on predictive performance. 

In a recursive, real-time decision-making exercise, we find substantial differences at various periods of time between the policy recommendations of BPDS and the traditional Bayesian model averaging approach, though good concordance at other times. When recommended policy decisions differ between the approaches, in most cases the BPDS policy paths are more intuitively sensible and less extreme than under BMA, and more consistent with the actual decisions made by the policymakers at the time. The case study presented investigates and interprets aspects of BPDS in terms of differential model weights based on historical information alone, and then updated based on identified optimal decisions, with consequent insights into how the differences relative to standard BMA arise and are exploited. This case study is a first step towards broader development and evaluation of BPDS in a setting with larger numbers of econometric models. Parallel next steps can naturally be expected include broader evaluation in terms of multiple models, and developments for scenario forecasting. Such extensions, in collaboration with policymakers, will advance understanding the sensitivity of model-based recommendations relative to chosen potential economic scenarios. 
 
\setstretch{1.1}
\section*{Acknowledgements} Research of Emily Tallman was partially supported by the US National Science Foundation through NSF Graduate Research
Fellowship Program grant DGE~2139754. Any opinions, findings, and conclusions or recommendations expressed in this material are those of the authors and do not necessarily reflect the views of the National Science Foundation. Further, the views expressed in this paper are solely those of the authors and may differ from the official views of the Bank of Canada. No responsibility for the views expressed in this paper should be attributed to the Bank of Canada.
 
\setstretch{0.9}\small
\bibliographystyle{Chicago} 
\bibliography{ChernisEtAlBPDS2024}
 
\setstretch{1}\normalsize
\section*{Appendix}
\subsection*{Appendix A: Evaluation of Tilting Vectors} 
Evaluation of tilting vectors $\btau(\x)$ involves Monte Carlo integration and numerical optimization, briefly summarized here. This follows and customizes the general results in~\citet[][section 4.4]{TallmanWest2023}. 

The vector $\btau(\x)$ is implicitly defined to satisfy  
$E_f[\s_j(\y,\x_j)|\x] = \m_f$, where $\m_f$ is the chosen target score and 
expectation is with respect to 
$$f(\y,\cM_j) = \tilde\pi_j(\x) f_j(\y|\x,\cM_j) = k(\x) \pi_j(\x) \alpha_j(\y|\x) p_j(\y|\x,\cM_j), \quad j=\seq 0J, $$ with components given in~\eqn{fj}. With $\alpha_j(\y|\x)$ of the exponential form in~\eqn{alphajBPDS}, 
numerical optimization   aims to solve 
\beq{solvefortau}
k(\x)\sum_{j=\seq 0J} \int_\y \pi_j(\x) \e^{\btau(\x)'\s_j(\y,\x_j)} p_j(\y|\x,\cM_j) d\y - \m_f = \bzero
\eeq
for $\btau(\x).$ The normalizing term $k(\x)$ is defined via $k(\x)^{-1} = \sum_{j=\seq 0J} \pi_j(\x)a_j(\x)$
where  
\beq{aj} a_j(\x) =  \int_\y  \e^{\btau(\x)'\s_j(\y,\x_j)}  p_j(\y|\x,\cM_j) d\y, \quad j=\seq 0J. \eeq
At any given $\x,$ the integrals in equations~(\ref{eq:solvefortau},\ref{eq:aj}) 
are approximated using Monte Carlo integration based on random samples from each of the $p_j(\y|\x,\cM_j).$ This gives the ingredients for numerical search over $\btau(\x)$ using derivative-free optimization methods such as those used in the current paper.   

For completeness, we note that the same approach applies where gradient-based optimization is considered. Newton-Raphson and allied methods will require the vector derivative $\dot a_j(\x)$ of $a_j(\x)$ with respect to $\btau(\x),$ and may also require the matrix of second derivatives $\ddot a_j(\x).$ These are given by
$$\begin{aligned} 
\dot a_j(\x)  &  =  \int_\y \s_j(\y, \x_j) \e^{\btau(\x)'\s_j(\y,\x_j)}  p_j(\y|\x, \cM_j) d\y,  \\
\ddot a_j(\x)  &  = \int_\y \s_j(\y, \x_j) \s_j(\y, \x_j)' \e^{\btau(\x)'\s_j(\y,\x_j)}  p_j(\y|\x, \cM_j) d\y.  
\end{aligned}$$
Hence the Monte Carlo integration approach used to evaluate the $a_j(\y)$ also delivers approximations to these derivatives.

\subsection*{Appendix B: Conditional Forecasts} 

This section discusses the construction of conditional forecasting in VAR models, following~\cite{chan-cond-forecast} and discussion in~\cite{antolindiaz-struct-2021}.

\subsubsection*{Observational Constraints}

The general theory for constraining on exact or uncertain linear constraints in multivariate normal forecast distributions~\citep[e.g.,][sect.~16.3.2]{WestHarrison1997} underlies these results used in that paper and here in the specific case of VAR models.  

The $n \times 1$ vector time series $\y_t = (y_{t1},\dots,y_{tn})'$ follows a  VAR$(p)$ in structural form 
\begin{equation}\label{eq:SVAR}
    \A_0\y_t = \a + \A_1y_{t-1} + \dots + \A_p y_{t-p} + \bvarepsilon_t, \quad \bvarepsilon_t \sim N(\bzero_n,\I_n)
\end{equation}
with the following terms: $\a$ is a $n \times 1$ vector of intercepts, $ \A_1,\dots,  \A_p$ are the $n \times n$ VAR coefficient matrices, $ \A_0$ is a full-rank contemporaneous impact matrix, $\bzero_n$ is a $n \times 1$ zero vector, and $\I_n$ is the $n\times n$ identity matrix.

Given the history of observations $\{\y_{1-p},\dots,\y_T\}$ at any time $T$, unconditional forecasting of the $n$ variables for the next $h$ periods, 
namely  the $nh\times 1$ vector $\y_{T+1:T+h} =  (\y'_{T+1},\dots,\y'_{T+h})'$, is based on the implied form
\begin{equation}\label{eq:stack_forecast}
    \H\y_{T+1:T+h} = \c + \bvarepsilon_{T+1:T+h}, \quad  \bvarepsilon_{T+1:T+h} \sim N(\bzero_{nh},\I_{nh})
\end{equation}
where
$$ 
    \c = 
    \begin{bmatrix}
        \a + \sum_{j=1}^{p} \A_j\y_{T+1-j}\\
        \a + \sum_{j=2}^{p} \A_j\y_{T+1-j}\\
        \vdots \\
        \a + \A_p\y_{T}\\
        \a \\
        \vdots \\
        \a       
    \end{bmatrix}, \ \ \ 
    \H = 
    \begin{bmatrix}
        \A_0 & \bzero_{n \times n} & \dots                          & \dots & \dots & \dots & \dots & \bzero_{n \times n} \\
        -\A_1 & \A_0           & \bzero_{n \times n}        & \dots & \dots & \dots & \dots & \bzero_{n \times n} \\
        -\A_2 & -\A_1          &\A_  0                  & \bzero_{n \times n} & \dots & \dots & \dots  & \bzero_{n \times n} \\
        \vdots         & \ddots                & \ddots                         & \ddots     & \dots& \dots& \dots& \vdots\\ 
        -\A_p  & \dots                  & \dots & -\A_1 & \A_0 & \bzero_{n \times n} & \dots & \bzero_{n \times n} \\
         \bzero_{n \times n} & \dots  & \dots & \ddots & \ddots & \ddots &\dots &\vdots\\         
         \vdots         & \dots                & \dots                         & \dots     & \ddots& \ddots& \ddots& \vdots\\ 
         \bzero_{n \times n} & \dots  & \bzero_{n \times n} & -\A_p  & \dots  & \dots & -\A_1 & \A_0\\         
    \end{bmatrix}
$$  
with $\bzero_{n \times n}$ the $n \times n$ zero matrix. Since $\A_0$ is full rank and $|\H| = |\A_0|^h \neq 0$, then $\H$ is non-singular and   
\begin{equation}\label{eq:uncond_forecast}
    \y_{T+1:T+h} \sim N(\m,\M) \quad \textrm{with \ moments} \quad \m = \H^{-1} \c \ \ \textrm{and}\ \ \M = (\H'\H)^{-1}.
\end{equation}
Now consider a set of $r<nh$ linear restrictions on the path of future observables, namely
\begin{equation} \label{eq:restr}
    \R\y_{T+1:T+h} \sim N(\r,\bOmega)
\end{equation}
where  $\R$ is a chosen $r \times nh$ constant matrix with full row rank,  $\r$ is an  $r \times 1$ vector mean, and  $\bOmega$  is the corresponding $r \times r$ variance matrix related to the restrictions.  This can be regarded as a set of uncertain constraints $\r$ when $\bOmega$ is non-zero, or exact constraints $\R \y_{T+1:T+h}=\r$ in the case $\bOmega = \bzero$.

Conditioning the forecast distribution of~\eqn{uncond_forecast} on this constraint information yields the updated distribution that follows; details follow from~\citet[sect.~16.3.2][]{WestHarrison1997} and \citet{chan-cond-forecast}. The updated distribution for the path is 
$$
    \y_{T+1:T+h} |\R,\r,\bOmega \sim N(\m^*,\M^*),
$$
with 
$$
\m^* = \m + \A(\r-\R\m)\quad\textrm{and}\quad \M^* =  \M+\A(\bOmega-\R\M\R')\A' \quad\textrm{where}\quad\A = \M\R'(\R\M\R')^{-1}.
$$  


The special case of exact constraints $\R \y_{T+1:T+h}=\r$ has constraint uncertainty matrix $\bOmega=\bzero$ so that $\M^*$ reduces to $\M^*= \M-\A\R\M\R'\A'.$ This is relevant in applications where it is justifiable to assume exact constraints. In our policy setting, the targeted constraint vector $\r$ is just that, a target policy path, for example, so there will always be some level of uncertainty. The starting point is 
uncertainty represented by the variance matrix $V(\R\y_{T+1:T+h}) = \R\M\R'.$ 
As argued in~\cite{antolindiaz-struct-2021}, adopting $\bOmega = \R\M\R'$ represents a position that admits relevant and {\em conservative} levels of such uncertainty. This is used in the application of our paper as we constrain on candidate values of the policy path in considering conditional forecasts, hedged with uncertainty. 

\subsubsection*{Constraints on Structural Shocks}

The development in the previous section implicitly involves constraints imposed on all of the structural shocks of the model. In conditional forecasting exercises, it is often necessary to focus on only specific structural shocks, set at restricted values (exact or uncertain), to obtain finer control. For example, a forecast conditional on an increasing policy rate path might result in an increase in inflation (i.e., the \lq\lq price puzzle") as the forecast is driven by reduced form shocks that are a mix of the structural shocks (i.e., demand, supply, monetary). For further discussion, see~\cite{antolindiaz-struct-2021}. Due to these issues, we impose structural restrictions so that changes in the policy rate are driven by monetary policy shocks (i.e., decisions from the Central Bank) rather than as reactions to other shocks. This can be done by considering the restrictions
\begin{equation}\label{eq:struct_restr}
    \W\bvarepsilon_{T+1:T+h} \sim N(\w,\bPsi)
\end{equation}
where $\W$ is a $w \times nh$ full rank-selection matrix, $\w$ is a $w \times 1$ vector of constants and $\bPsi$ is a $w\times w$ covariance matrix. 
Substituting~\eqn{stack_forecast} into~\eqn{struct_restr} for $\bvarepsilon_{T+1:T+h}$ 
results in 
\begin{equation}\label{eq:struct_restr2}
    \W\H\y_{T+1:T+h} \sim N(\W\c,\I_w)
\end{equation}
Then combining~\eqn{struct_restr2} with~\eqn{restr} yields
\begin{equation}\label{eq:struct_cond}
    \underbrace{\begin{bmatrix}
        \R_0\\
        \W\H
    \end{bmatrix}
    }_{\R}
\y_{T+1:T+h}  \sim N \Biggl(
    \underbrace{\begin{bmatrix}
        \r_0\\
        \W\c
    \end{bmatrix}
    }_{\r},
   \underbrace{\begin{bmatrix}
        \bOmega_0 & \bzero_{r_0 \times w}\\
        \bzero_{w \times r_0} & \I_w
    \end{bmatrix}}
    _{\bOmega}
    \Biggl)
\end{equation}
where $\R_0$ is a selection matrix for observable restrictions, $\r_0$ is a vector of restrictions on observables, and $\Omega_0$ is a matrix of covariance restrictions on observables. Thus, by setting $\R, \r$, and $\bOmega$ as indicated in~\eqn{struct_restr2}, the structural restrictions can be regarded as a specific case of ~\eqn{restr}.
In our applications, we set $\r_0 = \x$ and $\W = \bzero$ for all $\bvarepsilon_{T+1:T+h}$ except those associated with the monetary policy shock-- those are kept unrestricted. This amounts to conditioning on a proposed decision vector $\x$ and assuming it is driven only by monetary policy shocks. Other conditioning assumptions are possible. For example, in addition to assuming the driving shock is monetary policy, we could assume a vector of positive/negative demand shocks representing different future scenarios.

\end{document}